\documentclass[aps,prl,reprint,superscriptaddress]{revtex4-1}

\usepackage{hyperref}
\usepackage{graphicx}  
\usepackage{dcolumn}   
\usepackage{bm}        
\usepackage{amssymb}   
\usepackage{epstopdf}
\usepackage{amsmath}
\usepackage{xspace}
\usepackage{verbatim}
\usepackage{color}
\usepackage{soul}
\usepackage[export]{adjustbox}

\hypersetup{
    colorlinks=true,linkcolor=blue,citecolor=blue,
    filecolor=blue,urlcolor=blue,breaklinks=true
}

\begin{document}
  


\title{Global sensing and its impact for quantum many-body probes with criticality}

\author{Victor Montenegro}
\email{vmontenegro@uestc.edu.cn}
\affiliation{Institute of Fundamental and Frontier Sciences, University of Electronic Science and Technology of China, Chengdu 610051, PR China}

\author{Utkarsh Mishra}
\email{utkarsh.mishra@uestc.edu.cn}
\affiliation{Institute of Fundamental and Frontier Sciences, University of Electronic Science and Technology of China, Chengdu 610051, PR China}

\author{Abolfazl Bayat}
\email{abolfazl.bayat@uestc.edu.cn}
\affiliation{Institute of Fundamental and Frontier Sciences, University of Electronic Science and Technology of China, Chengdu 610051, PR China}

\date{\today}

\begin{abstract}
Quantum sensing is one of the key areas which exemplifies the superiority of quantum technologies. Nonetheless, most quantum sensing protocols operate efficiently only when the unknown parameters vary within a very narrow region, i.e., local sensing. Here, we provide a systematic formulation for quantifying the precision of a probe for multi-parameter global sensing when there is no prior information about the parameters. In many-body probes, in which extra tunable parameters exist, our protocol can tune the performance for harnessing the quantum criticality over arbitrarily large sensing intervals. For the single-parameter sensing, our protocol optimizes a control field such that an Ising probe is tuned to always operate around its criticality. This significantly enhances the performance of the probe even when the interval of interest is so large that the precision is bounded by the standard limit. For the multi-parameter case, our protocol optimizes the control fields such that the probe operates at the most efficient point along its critical line. Interestingly, for an Ising probe, it is predominantly determined by the longitudinal field. Finally, we show that even a simple magnetization measurement significantly benefits from our optimization and moderately delivers the theoretical precision.
\end{abstract}

\maketitle

\textit{Introduction.---} The emerging field of quantum sensing exploits quantum features for developing a new class of sensors with unprecedented precision~\cite{Degen-2017, Braun-2018, Toth-2014, Giovannetti-2011, Albarelli-2017, Giovannetti-2001}. Originally, the superiority of quantum sensors was shown by exploiting quantum superposition of GHZ-type states in non-interacting particles~\cite{Boto-2000, Leibfried-2004, Giovannetti-2006, Giovannetti-2004}. Such sensors use the resources (e.g., the number of particles $L$) more efficiently to enhance their precision, quantified by the variance of the estimation, from the usual classical standard limit (bounded by $1/L$) to the Heisenberg limit (bounded by $1/L^2$)~\cite{Belliardo-2020, Roy-2008}. However, if the particles interact, the precision goes down~\cite{boixo2007generalized,de2013quantum,skotiniotis2015quantum,pang2014quantum}. Moreover, the GHZ states are difficult to create and prone to decoherence~\cite{Albarelli-2018, Dur-2004, Demkowicz-2012, Matsuzaki-2011, Shaji-2007, Huelga-1997}. Hence, developing these types of sensors for many particles, where the quantum enhancement becomes significant, is extremely challenging in practice. To overcome such difficulties, a plethora of novel methods and systems have been exploited for sensing purposes, including quantum control techniques~\cite{cai2012quantum,liu2017quantum,liu2017control,sekatski2017quantum}, machine learning algorithms~\cite{hentschel2011efficient,xu2019generalizable,cimini2019calibration}, hybrid variational methods~\cite{meyer2020variational}, feedback schemes~\cite{lovett2013differential,yuan2015optimal,pang2017optimal,naghiloo2017achieving}, quantum chaos~\cite{fiderer2018quantum}, periodically driven systems~\cite{lang2015dynamical,mishra2020driving} and sequential measurements~\cite{bonato2016optimized,higgins2007entanglement,berry2009perform,jones2020remote}. 

Strongly correlated many-body systems are among the efficient quantum probes~\cite{Beau-2017, Guo-2016, Rey-2007, Choi-2008, Boixo-2008, Boixo-2009, Czajkowski-2019}. In particular, the ground state of many-body systems with quantum phase transitions is known to provide quantum-enhanced sensing at the vicinity of their critical point~\cite{Rams-2018, Invernizzi-2008, Zanardi-2008, Gammelmark-2011, Giulio-2014, Bina-2016, Frerot-2018, Chu-2021, Garbe-2020, Rossi-2017}. These schemes, unlike the GHZ-based quantum sensing protocols, truly exploit the interaction between the particles and because of operation at equilibrium they benefit from easier preparation and more robustness against decoherence. However, the quantum-enhanced sensing only occurs at the vicinity of the critical point~\cite{Rams-2018}, making these sensors most suitable for local sensing, where the unknown parameter varies within a very narrow interval. Hence, tuning the system to operate optimally at its quantum phase transition point can be very elusive and practically demanding, e.g., adaptive sensing strategies with complex measurement types have to be employed~\cite{Braun-2018, Mehboudi-2016,Okamoto-2012, Okamoto-2017, Fujiwara-2006, Fujiwara-2011, Wiseman-1995, Higgins-2007, Armen-2002}. A key open question is whether one can employ such sensors for global sensing, where the unknown parameter varies over a wide range. While in the particular case of temperature, there have been some efforts for the formulation of global thermometry~\cite{Bayat-2020, Correa-2020}, the problem is still open for general quantum sensing.
\begin{figure}[t]
\centering\includegraphics[width=\linewidth]{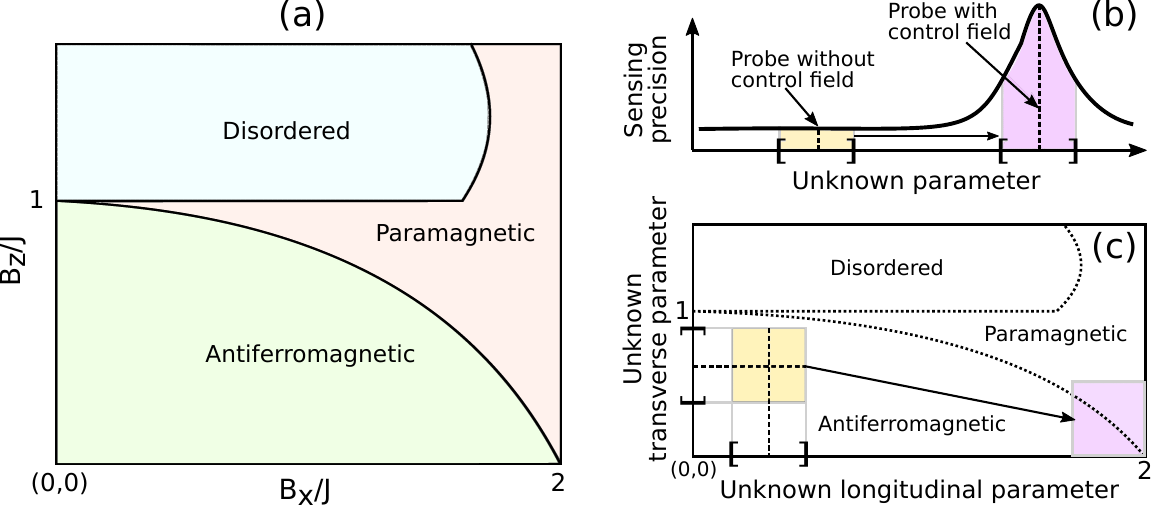}
\caption{(a) Schematic of the full phase diagram for an Ising model in the presence of a skew magnetic field (see Ref.~\cite{Bonfim-2019} for details). (b) By optimizing the control field, the probe is shifted in its phase diagram such that the sensing interval is located around the criticality, where the quantum Fisher information is maximum. (c) For multi-parameter sensing, the optimization of the probe also displaces the sensing area. However, the location depends on the size of the sensing area.}\label{fig:the-model}
\end{figure}

In this letter, we formulate a systematic approach for multi-parameter global sensing, where the unknown parameters can vary over arbitrarily large intervals. Our protocol applies to any sensing protocol and provides a systematic approach for optimizing the probe. In particular, for many-body sensors, we show that one can genuinely exploit the criticality as a resource for enhancing the global multi-directional magnetometry precision.

\textit{Parameter estimation.---}  Every sensing protocol contains three essential steps: (i) choosing an appropriate probe; (ii) gathering data through repeatedly performing specific types of measurements on the probe; and (iii) feed the gathered data into an estimator to infer the value of the unknown parameters.
The precision of the estimation of an unknown parameter $\boldsymbol{h}{=}(h_1, h_2, \ldots)$ obeys the Cram\'{e}r-Rao inequality~\cite{Cramer-1946, Braunstein-1994}
\begin{equation}
\mathrm{Cov}[\boldsymbol{h}] \geq M^{-1} \mathrm{inv}\left[\mathcal{F}_C(\boldsymbol{h})\right] \geq M^{-1} \mathrm{inv}\left[\mathcal{F}_Q(\boldsymbol{h})\right],\label{eq:multi-QCRB}
\end{equation}
where $\mathrm{Cov}[\boldsymbol{h}]$ is the covariance matrix whose elements are $[\mathrm{Cov}[\boldsymbol{h}]]_{\mu, \nu}{=}\langle h_\mu h_\nu\rangle{-}\langle h_\mu \rangle \langle h_\nu \rangle$, $M$ is the total number of measurements, and $\mathcal{F}_C(\boldsymbol{h})$ and $\mathcal{F}_Q(\boldsymbol{h})$ are the classical and quantum Fisher information (QFI) matrices, respectively~\cite{Paris-2009}. For a given quantum probe, with density matrix $\rho(\boldsymbol{h})$, and a specific POVM measurement $\{\Pi_k\}$, the bound is given by the classical Fisher information matrix $[\mathcal{F}_C(\boldsymbol{h})]_{\mu, \nu}{=}\sum_{k} p_k(\boldsymbol{h}) \left[\partial_\mu \log p_k(\boldsymbol{h})\right]\left[\partial_\nu \log p_k(\boldsymbol{h})\right]$,
where $p_k(\boldsymbol{h}){=}\mathrm{Tr}[\rho(\boldsymbol{h})\Pi_k]$ is the probability of measurement outcome $k$, and $\partial_\nu{:=}\partial/\partial h_\nu$. By optimizing over all possible POVMs, one can tightens the bound to be given by the QFI $\mathcal{F}_Q(\boldsymbol{h})$~\cite{Paris-2009, Braunstein-1994}, which can be simplified for pure states $\rho(\boldsymbol{h}){=}|\Phi(\boldsymbol{h})\rangle\langle \Phi(\boldsymbol{h})|$ as $[\mathcal{F}_Q(\boldsymbol{h})]_{\mu,\nu}{=}4\mathrm{Re}\left[\langle \partial_\mu \Phi | \partial_\nu \Phi\rangle{+}\langle \Phi | \partial_\mu \Phi\rangle\langle \Phi | \partial_\nu \Phi\rangle\right]$~\cite{Liu-2019}. In the case of single parameter, the Eq.~\eqref{eq:multi-QCRB} reduces to
\begin{equation}
\delta h^2 \geq M^{-1} \mathcal{F}_Q(h)^{-1},\label{eq:uncertainty-single}
\end{equation}
in which the QFI provides the ultimate bound for the variance of the estimation. Note that, for any given probe one can indeed saturate the inequality of Eq.~\eqref{eq:uncertainty-single} by using an appropriate measurement setup computed through symmetric logarithmic derivatives~\cite{Paris-2009, Seveso-2020, Albarelli-2020} and an optimal estimator (which for large data set is known to be Bayesian~\cite{LeCam-1986, Hradil-1996, Pezze-2007, Rubio-2019, Olivares-2009}). However, the optimal measurement basis, in general, depend on the unknown parameter $h$ which varies over an interval $[h^{\mathrm{min}}, h^{\mathrm{max}}]$. The optimal sensing is thus applicable only when $\Delta h {=} h^{\mathrm{max}} {-} h^{\mathrm{min}}$ is small, which is called \textit{local} sensing. In this situation, the optimal measurement can be chosen for $h^\mathrm{cen}{=}(h^{\mathrm{max}}{+}h^{\mathrm{min}})/2$. For large $\Delta h$, the criteria for the best sensing protocol is indeed unknown. The situation for multi-parameter estimation is more complex as even for local sensing due to the non-commutativity of the optimal POVMs for different parameters, the Cram\'{e}r-Rao bound may not be achievable~\cite{Albarelli-2020}.

\textit{Single parameter global sensing.---} We first introduce the concept of global sensing for a single-parameter estimation, and later in the paper, we generalize it for multi-parameter cases. As mentioned before, in the case of local sensing (i.e., small $\Delta h$), the Cram\'{e}r-Rao bound in Eq.~\eqref{eq:uncertainty-single} can always be saturated. Nonetheless, the sensing procedure might still be highly sub-optimal due to the bad choice of the probe. Hence, an optimal local sensing algorithm requires optimization of $\mathcal{F}_Q(h^\mathrm{cen})$ with respect to the parameters of the probe. For large $\Delta h$, which is called \textit{global} sensing, the situation is more complex as: (i) in general, the optimal measurement basis varies over $\Delta h$ and no measurement setup can saturate the Cram\'{e}r-Rao bound over the entire interval; and (ii) it is not clear which quantity has to be optimized to find the optimal probe. In the following, we address this problem.

To formulate the global sensing, we first quantify the average uncertainty of the estimation via $\int_{\Delta h} \delta h^2 f(h) dh$, where $f(h)$ is the prior information, i.e. the probability distribution, of the unknown parameter $h$ over the sensing interval $\Delta h$ of interest. From Eq.~\eqref{eq:uncertainty-single}, one can easily show that this average uncertainty is bounded by
\begin{equation}
g(\boldsymbol{B}) := \int_{\Delta h} \frac{f(h)}{\mathcal{F}_Q(h|\boldsymbol{B})}dh,\label{eq:g-single}
\end{equation}
where $\boldsymbol{B}{=}(B_1,B_2,\ldots)$ are external tunable parameters interacting with the probe. We define the minimization of $g(\boldsymbol{B})$ with respect to control parameters $\boldsymbol{B}$ as a figure of merit for finding the optimal probe, namely $g(\boldsymbol{B}^*){:=}\mathrm{min}_{\boldsymbol{B}} \left[ g(\boldsymbol{B}) \right]$. Throughout this paper, we assume no prior information about the unknown parameter $h$, and thus, $f(h)$ takes a uniform distribution, namely $f(h){=}1/\Delta h$. For local sensing, i.e. small $\Delta h$, the QFI is almost constant over the interval, and thus, $g(\boldsymbol{B}){\approx}1/\mathcal{F}_Q(h^\mathrm{cen}|\boldsymbol{B})$. Therefore, the minimization of $g(\boldsymbol{B})$ is equivalent to maximize $\mathcal{F}_Q(h^\mathrm{cen}|\boldsymbol{B})$.

\textit{Many-body probe for magnetometry.---} To illustrate the relevance of our general formulation for global sensing, which it can be applied to any sensing protocol independent of the choice of the probe, we exploit a chain of $L$ interacting spin$-1/2$ particles with Ising Hamiltonian to sense a random static magnetic field. For simplicity and without loss of generality, we assume that the $y$ component of the magnetic field is zero. The Hamiltonian is
\begin{equation}
H = J\sum_{i=1}^{L} \sigma_x^i \sigma_x^{i+1} - \sum_{i=1}^{L}[(B_x + h_x)\sigma_x^i + (B_z + h_z)\sigma_z^i],\label{eq:hamiltonian}
\end{equation}
where, $\sigma_\alpha^i$ ($\alpha{=}x,z$) is the Pauli operator at site $i$, $J>0$ is the exchange interaction, $\boldsymbol{B}{=}(B_x, B_z)$ is the control magnetic field which can be tuned, $\boldsymbol{h}{=}(h_x, h_z)$ is the random field to estimate, and periodic boundary condition is imposed. The ground state $|\Phi\rangle$, for which the phase diagram is shown in Fig.~\ref{fig:the-model}(a), can be used for detecting the unknown field $\boldsymbol{h}$. See Refs.~\cite{Venuti-2007, Zanardi-2006, Zhou-2008, Zanardi-2007-crit, Rezakhani-2010} for the link between the characterization of quantum phase transition via metric tensors. The phase diagram is uniquely determined by $(h_z{+}B_z)/J$ and $(h_x{+}B_x)/J$. Thus, by tuning $\boldsymbol{B}$ one can shift the phase diagram for the unknown parameter $\boldsymbol{h}$. For example, in the absence of longitudinal field (i.e. $B_x{=}h_x{=}0$), the ground state $|\Phi\rangle$ can be solved analytically using Jordan-Wigner transformation and is known to have a quantum phase transition at $h^{\mathrm{crit}}{:=}B_z{+}h_z{=}\pm J$~\cite{Fisher-1995, Sachdev-2011}, which leads to an enhanced QFI with scaling ${\sim}L^2$~\cite{Rams-2018, Zanardi-2008} (see the Supplementary Material for details). Interestingly, away from criticality, the QFI scales as ${\sim}L\xi^{-1}$, where $\xi {\sim}|(B_z{+}h_z){-}h^\mathrm{crit}|^{-1}$ is the correlation length~\cite{Rams-2018}. Since the quantum-enhanced sensing is lost when the probe operates away from criticality, our general formulation for global sensing stands as the most suitable for these type of sensors.

\textit{Example 1: Transverse field Ising probe.---} In this section, we use the Ising many-body probe, given in Eq.~\eqref{eq:hamiltonian}, for single-parameter sensing when only transverse field exists, namely $B_x{=}h_x{=}0$.
\begin{figure}[t]
\centering \includegraphics[width=\linewidth]{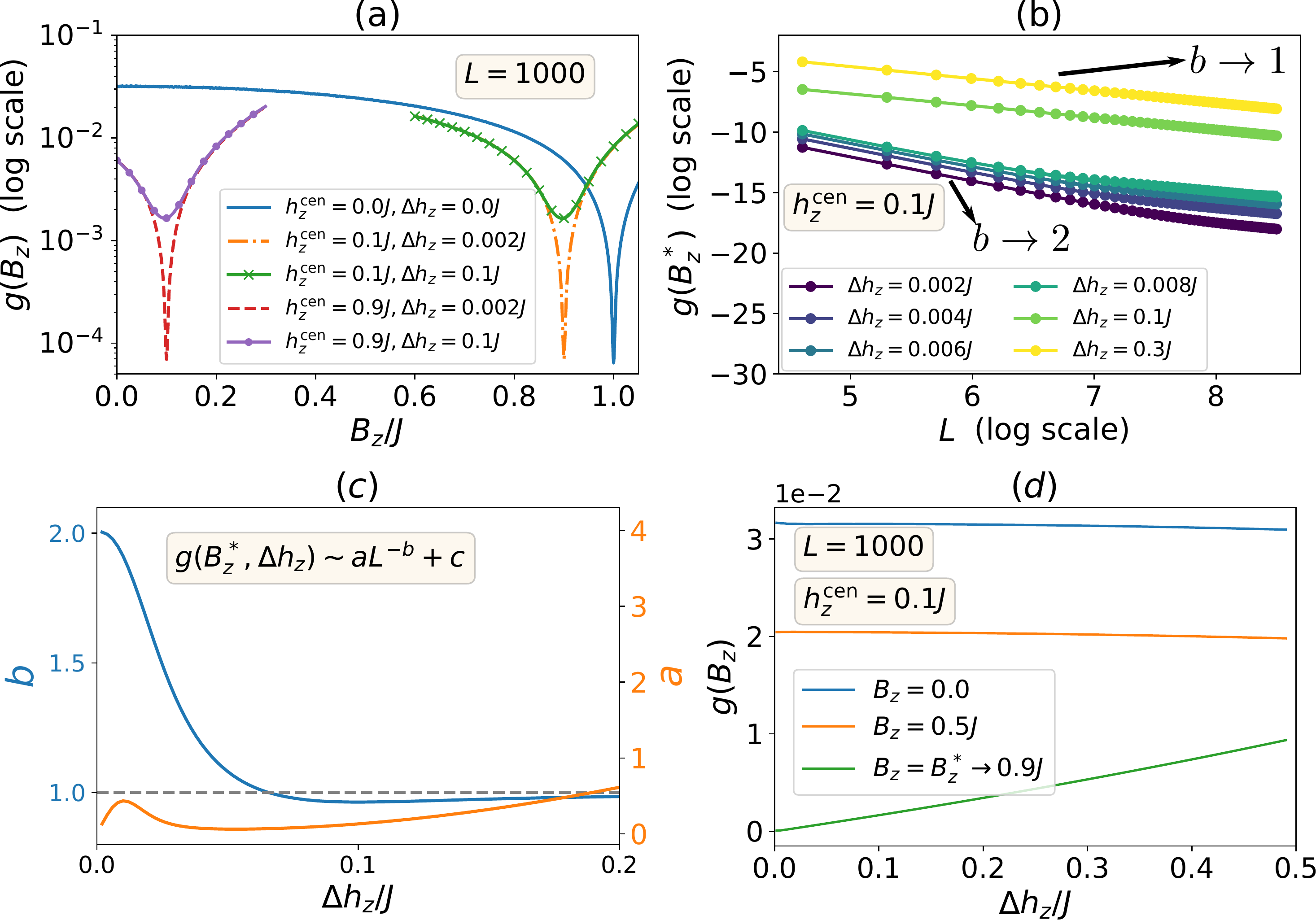}
\caption{(a) Average uncertainty $g(B_z)$ as function of $B_z/J$ for different values of interval widths $\Delta h_z$ and centers $h_z^\mathrm{cen}$. The optimal probe can always be found by tuning the control field such that $g(B_z^*){=}\mathrm{min}_{B_z}\left[ g(B_z) \right]$. (b) $g(B_z^*)$ (shown by markers) and its corresponding fitting function $aL^{-b}+c$ (solid lines) as a function of $L$ for various choices of $\Delta h_z$. (c) Fitting coefficients $a$ and $b$ versus the controlled field $B_z/J$. (d) $g(B_z)$ is plotted as a function of $\Delta h_z/J$ for various choices of $B_z$.}\label{fig:figure_1}
\end{figure}
To see the performance of this probe for global sensing, we numerically evaluate $g(B_z)$ for various $\Delta h$ for a system size $L{=}1000$. In Fig.~\ref{fig:figure_1}(a), we plot the global sensing performance $g(B_z)$ as function of $B_z/J$ for different values of interval  widths $\Delta h_z$ and centers $h_z^\mathrm{cen}$. For every $h_z^\mathrm{cen}$ and $\Delta h_z$, the average uncertainty $g(B_z)$ has always a minimum which takes place at a particular $B_z{=}B_z^*$, showing that one can always make the probe optimal by this choice of control field. Interestingly, the minimum value of $g(B_z)$ is independent of $h_z^\mathrm{cen}$, and is only determined by $\Delta h_z$. On the other hand, the optimal control field $B_z^*$ is almost independent of $\Delta h_z$ and only depends on $h_z^\mathrm{cen}$, such that $h_z^\mathrm{cen}{+}B_z^* \approx h^{\mathrm{crit}}$. This means that, the control field tends to shift the probe in its phase diagram such that the interval of sensing is located almost symmetrically around the critical point. This has been shown schematically in Fig.~\ref{fig:the-model}(b). To determine how the average uncertainty scales for the optimal probe, in Fig.~\ref{fig:figure_1}(b), we plot $g(B_z^*)$ as a function of $L$ for various choices of $\Delta h_z$. For small $\Delta h_z$, the average uncertainty scales as $g(B_z^*){\sim}1/L^2$, which is expected for local sensing. Remarkably, by increasing $\Delta h_z$, the scaling goes towards the standard limit, namely $g(B_z^*){\sim}1/L$. To quantify the transition from quantum enhanced sensing to the standard limit, we fit $g(B_z^*)$ with a function of the form $aL^{-b}+c$, $c{\rightarrow}0$. In Fig.~\ref{fig:figure_1}(c), we plot the fitting coefficients $a$ and $b$ as a function of the width $\Delta h_z/J$. As seen from the figure, when $\Delta h_z{\rightarrow}0$, one recovers the Heisenberg scaling, $g(B_z^*){\sim}{F}_Q^{-1}{\sim}1/L^2$, from the local sensing strategy. The quantum-enhanced sensing is captured by $b>1$ for which the precision surpasses the standard limit $b=1$. Remarkably, the region of quantum enhanced sensing is extended only until $\Delta h_z{\leq}0.07J$, beyond which the standard limit is restored. However, it is crucial to note that the optimization of the probe is still beneficial for sensing even though there is no quantum-enhanced advantage in the scaling. This can be seen in Fig.~\ref{fig:figure_1}(d) where $g(B_z)$ is plotted as a function of $\Delta h_z/J$ for various choices of $B_z$. As the figure shows, by the optimal choice of $B_z{=}B_z^*$ the average uncertainty remains lower than non-optimal values of the control field for all values of $\Delta h_z$. In other words, for large $\Delta h_z$, while $b{=}1$, the $a$ coefficient becomes smaller by optimizing $B_z$.

\textit{Multi-parameter global sensing.---} Thanks to the above analysis, one can readily generalize the global sensing for the multi-parameter case. To set the performance of different multi-parameter estimators, we recast the matrix bounds in Eq.~\eqref{eq:multi-QCRB} into scalar bounds. To do so, we introduce a (positive and real) weight matrix $\mathcal{W}$ such that~\cite{Albarelli-2020}
\begin{equation}
\mathrm{Tr}\left[\mathcal{W} \mathrm{Cov}[\boldsymbol{h}]\right] \geq M^{-1} \mathrm{Tr}\left[\mathrm{inv}\left[\mathcal{F}_Q(\boldsymbol{h}|\boldsymbol{B})\right] \mathcal{W} \right].\label{eq:scalar-bound}
\end{equation}
Throughout this work, we consider $\mathcal{W}$ to be identity, namely $\mathcal{W}{=}\mathbb{I}$. This $\mathcal{W}$ choice makes the left-hand side of the above inequality to be the sum of the variances of the unknown parameters. Inspired by the single-parameter case, we define the average uncertainty for the multi-parameter case as
\begin{equation}
g(\boldsymbol{B}) := \int_{\Delta \boldsymbol{h}} f(\boldsymbol{h}) \mathrm{Tr}[\mathrm{inv}[\mathcal{F}_Q(\boldsymbol{h}|\boldsymbol{B})]] d\boldsymbol{h},\label{eq:g-multi}
\end{equation}
where $f(\boldsymbol{h})$ is the prior probability distributions, which is assumed to be uniform, for magnetic field $\boldsymbol{h}$, and the integration is performed over the volume of all parameters. For the single-parameter case, the above equation reduces to Eq.~\eqref{eq:g-single}. To optimize the probe, one has to minimize $g(\boldsymbol{B})$ with respect to $\boldsymbol{B}$, i.e. $g(\boldsymbol{B}^*){:=} \mathrm{min}_{\boldsymbol{B}}[g(\boldsymbol{B})]$.

\textit{Example 2: Skew field Ising probe.---} In this section, we consider the probe of Eq.~\eqref{eq:hamiltonian}, when both $h_x$ and $h_z$ are sensed together. Since the system is not solvable anymore, we are restricted to short chains and exact diagonalization. Unlike the transverse field Ising model for which the criticality happens exactly at one point, here, there is a line of criticality in the plane of $(h_x, h_z)$\cite{Bonfim-2019}, see Fig.~\ref{fig:the-model}(a). Thus, the optimization of the probe is highly non-trivial as the control fields $(B_x, B_z)$ can shift the phase of the probe to operate anywhere along the critical line. For the case of local sensing, in which $\Delta h_x$ and $\Delta h_z$ are very small, the minimization of $g(\boldsymbol{B})$ reduces to the maximization of $\mathrm{Tr}[\mathrm{inv}[\mathcal{F}_Q( (h_x^{\mathrm{cen}}, h_z^{\mathrm{cen}})|\boldsymbol{B})]]$.

\begin{figure}[t]
\centering \includegraphics[width=\linewidth]{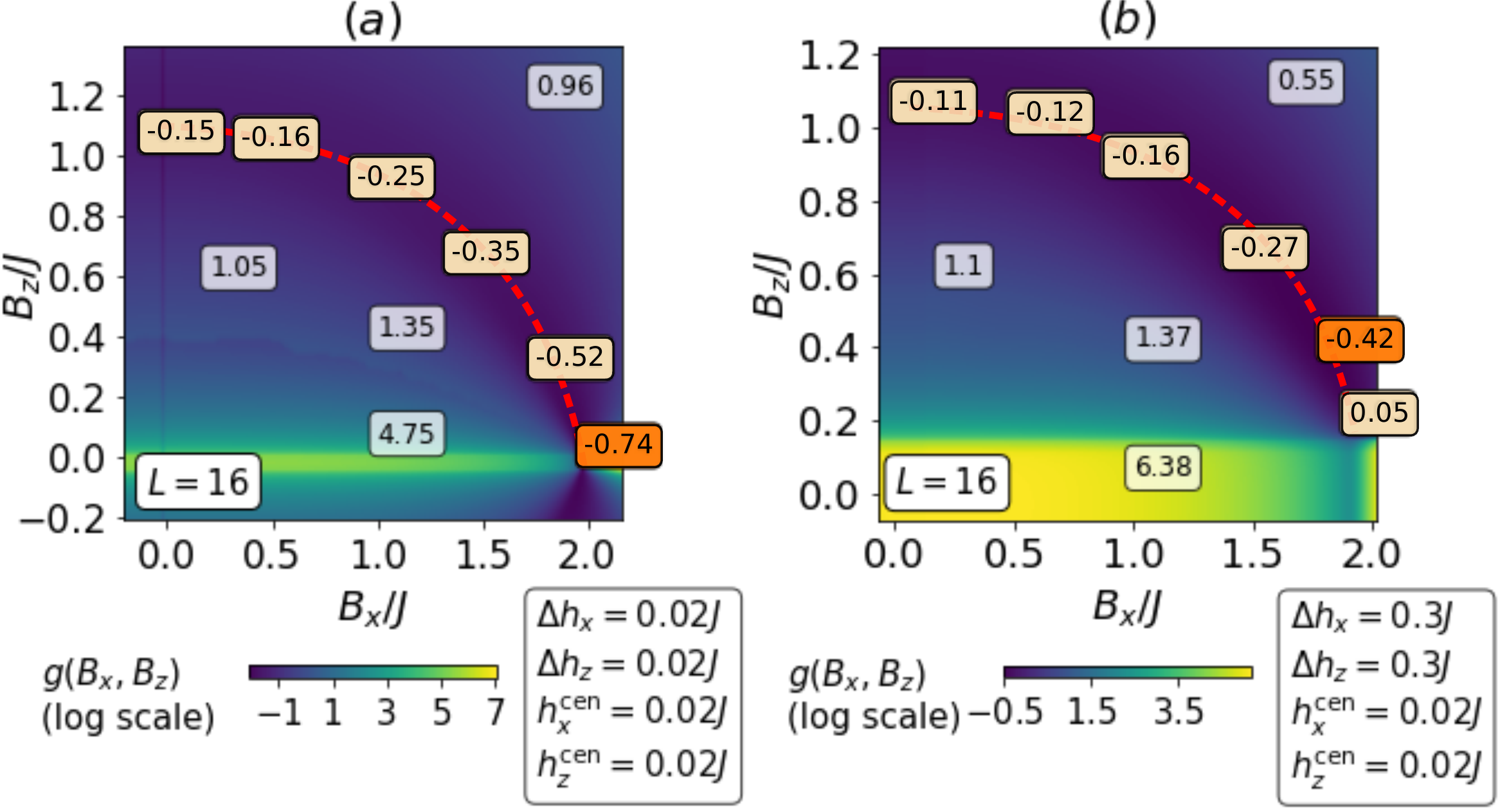}
    \caption{(a) $g(B_x,B_z)$ as a function of control fields $(B_x, B_z)$ for the case of an almost local sensing $\Delta h_x{=}\Delta h_z{=}0.02J$. (b) By increasing the widths to $\Delta h_x{=}\Delta h_z{=}0.3J$, one finds a non-trivial optimal probe along its critical line. The total system is $L = 16$ and centers are chosen to be $h_x^\mathrm{cen}{=}h_z^\mathrm{cen}{=}0.02J$. The dashed red lines represent the critical line and the numbers show the value of $g(B_x,B_z)$ at that point in the phase diagram. The global minimum is depicted in orange.}\label{fig:figure_2}
\end{figure}

In contrast to the single-parameter case, our analysis shows that, the optimal control fields not only depend on the location of $(h_x^\mathrm{cen}, h_z^\mathrm{cen})$, but also change by the choice of the widths $(\Delta h_x, \Delta h_z)$. Interestingly, the probe is always shifted somewhere near the critical line which is predominantly controlled by the longitudinal field. To have a quantitative analysis, in Fig.~\ref{fig:figure_2}(a), we plot $g(B_x,B_z)$ as a function of control fields for the case of an almost local sensing with $h_x^\mathrm{cen}{=}h_z^\mathrm{cen} {=}0.02J$ and the widths of $\Delta h_x{=}\Delta h_z{=} 0.02J$. As the figure shows, the optimal control fields are given by $B_x^*{=}1.98J $ and $B_z^*{=}-0.02J$. In Fig.~\ref{fig:figure_2}(b), we increase the width to $\Delta h_x{=}\Delta h_z{=}0.3J$ (i.e. global sensing). Interestingly, the optimal control fields shift to $B_x^*{=}1.8J$ and $B_z^*{=}0.4J$, which are different from the previous case. This shows that, the optimization of the probe for multi-parameter sensing is non-trivial and the optimal control fields depend on the sensing range. This is schematically explained in Fig.~\ref{fig:the-model}(c).

\begin{figure}[t]
\centering \includegraphics[width=\linewidth]{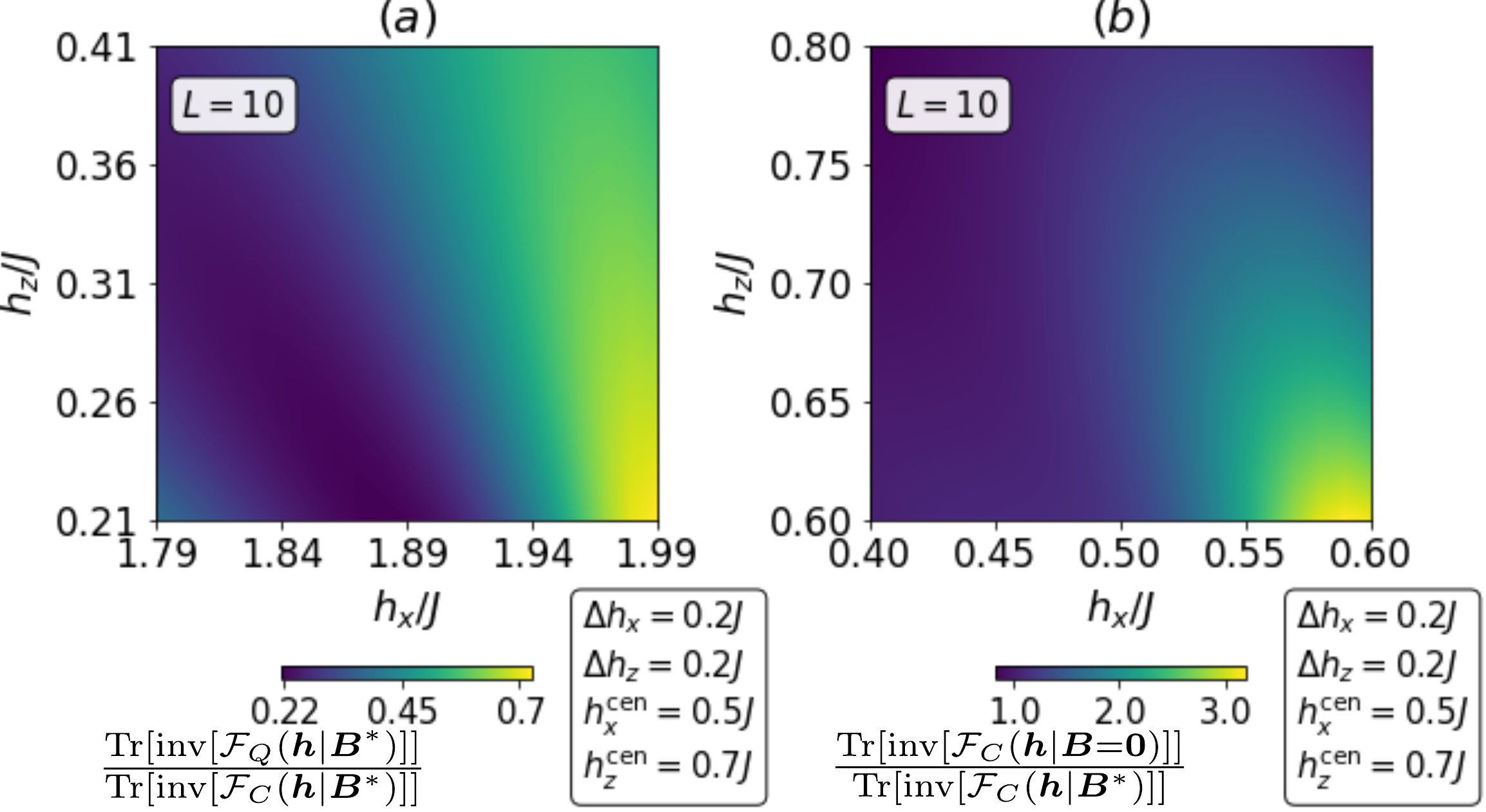}
\caption{(a) Efficiency of the magnetization measurement $\mathrm{Tr}[\mathrm{inv}[\mathcal{F}_Q(\boldsymbol{h}|\boldsymbol{B}^*)]]/\mathrm{Tr}[\mathrm{inv}[\mathcal{F}_C(\boldsymbol{h}|\boldsymbol{B}^*)]]$ as a function of $h_x/J$ and $h_z/J$. One can achieve the ultimate precision between $22\%-70\%$ for all values of $h_x/J$ and $h_z/J$. (b) Efficiency comparison between the optimal probe against the non-optimal one $\mathrm{Tr}[\mathrm{inv}[\mathcal{F}_C(\boldsymbol{h}|\boldsymbol{B}=\boldsymbol{0})]]/\mathrm{Tr}[\mathrm{inv}[\mathcal{F}_C(\boldsymbol{h}|\boldsymbol{B}^*)]]$ as a function of $h_x/J$ and $h_z/J$. The performance for the optimal probe exceeds considerably the non-optimal one. Other values are: $L{=}10$, $h_x^\mathrm{cen}{=}0.5J, h_z^\mathrm{cen}{=}0.7J$, and $\Delta h_x{=}\Delta h_z{=}0.2J$}\label{fig:figure_3}
\end{figure}

\textit{Sensing protocol.---} While our formulation for global sensing systematically provides a bound for the average uncertainty, it is not obvious whether this bound can be saturated, as no measurement basis can be optimal throughout the whole region. Here, we choose a simple measurement basis, namely magnetization $M{=}\sum_i \sigma_z^i$, which is independent of the unknown parameters. A measurement already available in ion traps~\cite{Kokail-2019} and superconducting quantum devices~\cite{Chiaro-2020, Arute-2019}. Although, this choice of measurement basis is not necessarily optimal, it has practical advantages due to its simplicity. In fact, measuring the ground state results in $L{+}1$ outcomes for which one can compute the classical Fisher information matrix. We consider a two parameter sensing with $h_x^\mathrm{cen}{=}0.5J, h_z^\mathrm{cen}{=}0.7J$, and $\Delta h_x{=}\Delta h_z{=}0.2J$. One can optimize the probe of Eq.~\eqref{eq:hamiltonian} using the algorithm above, i.e. minimizing $g(B_x, B_z)$, which results in $B_x^*{=}1.39J$ and $B_z^*{=}{-}0.39J$. Using this optimal probe, we perform the magnetization measurement and compute $\mathrm{Tr}[\mathrm{inv}[\mathcal{F}_C]]$ over the whole interval. To quantify the efficiency of this measurement, in Fig.~\ref{fig:figure_3}(a), we plot $\mathrm{Tr}[\mathrm{inv}[\mathcal{F}_Q(\boldsymbol{h}|\boldsymbol{B}^*)]]/\mathrm{Tr}[\mathrm{inv}[\mathcal{F}_C(\boldsymbol{h}|\boldsymbol{B}^*)]]$ as a function of $h_x/J$ and $h_z/J$ to compare the obtainable precision with the ultimate bound. Interestingly, despite using this simple measurement, the ultimate precision bound ranges between $22\%{-}70\%$ for all values of $h_x/J$ and $h_z/J$. To see the importance of optimizing the probe, in Fig.~\ref{fig:figure_3}(b), we plot the $\mathrm{Tr}[\mathrm{inv}[\mathcal{F}_C(\boldsymbol{h}|\boldsymbol{B}{=}\boldsymbol{0}))]]/\mathrm{Tr}[\mathrm{inv}[\mathcal{F}_C(\boldsymbol{h}|\boldsymbol{B}^*]]$ as a function of $h_x/J$ and $h_z/J$ to compare the performance of the magnetization measurement for the optimal versus the non-optimal probe, given by $B_x{=}0, B_z{=}0$. As seen from the figure, higher performance can be achieved by using the optimal probe. This shows that our procedure for global sensing can optimize the probe such that even with a simple measurement, one can harness the criticality to improve the precision significantly.

\textit{Conclusions.---} We present a formulation for multi-parameter global sensing which not only provides a bound for the average uncertainty, but also allows for systematic optimization of the probe. By applying our protocol to an Ising many-body probe, we show that one can indeed tune external control fields to harness the criticality for enhancing the sensing precision, even when the intervals of interests are so large that the Heisenberg limit is absent. While the optimal measurement basis remains an open problem, we show that a simple magnetization measurement can hugely benefit from our optimization.

\textit{Acknowledgments.---} This work is supported by the National Key R$\&$D program of China (Grant No. 2018YFA0306703) and National Science Foundation of China (Grants No. 12050410253, No. 92065115, and No. 12050410251). U.M. acknowledges funding from the Chinese Postdoctoral Science Fund for Grant No. 2018M643437. V.M. thanks the Chinese Postdoctoral Science Fund for Grant No. 2018M643435.

\bibliography{Magnetometry}

\begin{thebibliography}{92}%
\makeatletter
\providecommand \@ifxundefined [1]{%
 \@ifx{#1\undefined}
}%
\providecommand \@ifnum [1]{%
 \ifnum #1\expandafter \@firstoftwo
 \else \expandafter \@secondoftwo
 \fi
}%
\providecommand \@ifx [1]{%
 \ifx #1\expandafter \@firstoftwo
 \else \expandafter \@secondoftwo
 \fi
}%
\providecommand \natexlab [1]{#1}%
\providecommand \enquote  [1]{``#1''}%
\providecommand \bibnamefont  [1]{#1}%
\providecommand \bibfnamefont [1]{#1}%
\providecommand \citenamefont [1]{#1}%
\providecommand \href@noop [0]{\@secondoftwo}%
\providecommand \href [0]{\begingroup \@sanitize@url \@href}%
\providecommand \@href[1]{\@@startlink{#1}\@@href}%
\providecommand \@@href[1]{\endgroup#1\@@endlink}%
\providecommand \@sanitize@url [0]{\catcode `\\12\catcode `\$12\catcode
  `\&12\catcode `\#12\catcode `\^12\catcode `\_12\catcode `\%12\relax}%
\providecommand \@@startlink[1]{}%
\providecommand \@@endlink[0]{}%
\providecommand \url  [0]{\begingroup\@sanitize@url \@url }%
\providecommand \@url [1]{\endgroup\@href {#1}{\urlprefix }}%
\providecommand \urlprefix  [0]{URL }%
\providecommand \Eprint [0]{\href }%
\providecommand \doibase [0]{http://dx.doi.org/}%
\providecommand \selectlanguage [0]{\@gobble}%
\providecommand \bibinfo  [0]{\@secondoftwo}%
\providecommand \bibfield  [0]{\@secondoftwo}%
\providecommand \translation [1]{[#1]}%
\providecommand \BibitemOpen [0]{}%
\providecommand \bibitemStop [0]{}%
\providecommand \bibitemNoStop [0]{.\EOS\space}%
\providecommand \EOS [0]{\spacefactor3000\relax}%
\providecommand \BibitemShut  [1]{\csname bibitem#1\endcsname}%
\let\auto@bib@innerbib\@empty
\bibitem [{\citenamefont {Degen}\ \emph {et~al.}(2017)\citenamefont {Degen},
  \citenamefont {Reinhard},\ and\ \citenamefont {Cappellaro}}]{Degen-2017}%
  \BibitemOpen
  \bibfield  {author} {\bibinfo {author} {\bibfnamefont {C.~L.}\ \bibnamefont
  {Degen}}, \bibinfo {author} {\bibfnamefont {F.}~\bibnamefont {Reinhard}}, \
  and\ \bibinfo {author} {\bibfnamefont {P.}~\bibnamefont {Cappellaro}},\
  }\href {\doibase 10.1103/RevModPhys.89.035002} {\bibfield  {journal}
  {\bibinfo  {journal} {Rev. Mod. Phys.}\ }\textbf {\bibinfo {volume} {89}},\
  \bibinfo {pages} {035002} (\bibinfo {year} {2017})}\BibitemShut {NoStop}%
\bibitem [{\citenamefont {Braun}\ \emph {et~al.}(2018)\citenamefont {Braun},
  \citenamefont {Adesso}, \citenamefont {Benatti}, \citenamefont {Floreanini},
  \citenamefont {Marzolino}, \citenamefont {Mitchell},\ and\ \citenamefont
  {Pirandola}}]{Braun-2018}%
  \BibitemOpen
  \bibfield  {author} {\bibinfo {author} {\bibfnamefont {D.}~\bibnamefont
  {Braun}}, \bibinfo {author} {\bibfnamefont {G.}~\bibnamefont {Adesso}},
  \bibinfo {author} {\bibfnamefont {F.}~\bibnamefont {Benatti}}, \bibinfo
  {author} {\bibfnamefont {R.}~\bibnamefont {Floreanini}}, \bibinfo {author}
  {\bibfnamefont {U.}~\bibnamefont {Marzolino}}, \bibinfo {author}
  {\bibfnamefont {M.~W.}\ \bibnamefont {Mitchell}}, \ and\ \bibinfo {author}
  {\bibfnamefont {S.}~\bibnamefont {Pirandola}},\ }\href {\doibase
  10.1103/RevModPhys.90.035006} {\bibfield  {journal} {\bibinfo  {journal}
  {Rev. Mod. Phys.}\ }\textbf {\bibinfo {volume} {90}},\ \bibinfo {pages}
  {035006} (\bibinfo {year} {2018})}\BibitemShut {NoStop}%
\bibitem [{\citenamefont {T{\'{o}}th}\ and\ \citenamefont
  {Apellaniz}(2014)}]{Toth-2014}%
  \BibitemOpen
  \bibfield  {author} {\bibinfo {author} {\bibfnamefont {G.}~\bibnamefont
  {T{\'{o}}th}}\ and\ \bibinfo {author} {\bibfnamefont {I.}~\bibnamefont
  {Apellaniz}},\ }\href {\doibase 10.1088/1751-8113/47/42/424006} {\bibfield
  {journal} {\bibinfo  {journal} {Journal of Physics A: Mathematical and
  Theoretical}\ }\textbf {\bibinfo {volume} {47}},\ \bibinfo {pages} {424006}
  (\bibinfo {year} {2014})}\BibitemShut {NoStop}%
\bibitem [{\citenamefont {Giovannetti}\ \emph {et~al.}(2011)\citenamefont
  {Giovannetti}, \citenamefont {Lloyd},\ and\ \citenamefont
  {Maccone}}]{Giovannetti-2011}%
  \BibitemOpen
  \bibfield  {author} {\bibinfo {author} {\bibfnamefont {V.}~\bibnamefont
  {Giovannetti}}, \bibinfo {author} {\bibfnamefont {S.}~\bibnamefont {Lloyd}},
  \ and\ \bibinfo {author} {\bibfnamefont {L.}~\bibnamefont {Maccone}},\ }\href
  {\doibase 10.1038/nphoton.2011.35} {\bibfield  {journal} {\bibinfo  {journal}
  {Nature Photonics}\ }\textbf {\bibinfo {volume} {5}},\ \bibinfo {pages} {222}
  (\bibinfo {year} {2011})}\BibitemShut {NoStop}%
\bibitem [{\citenamefont {Albarelli}\ \emph {et~al.}(2017)\citenamefont
  {Albarelli}, \citenamefont {Rossi}, \citenamefont {Paris},\ and\
  \citenamefont {Genoni}}]{Albarelli-2017}%
  \BibitemOpen
  \bibfield  {author} {\bibinfo {author} {\bibfnamefont {F.}~\bibnamefont
  {Albarelli}}, \bibinfo {author} {\bibfnamefont {M.~A.~C.}\ \bibnamefont
  {Rossi}}, \bibinfo {author} {\bibfnamefont {M.~G.~A.}\ \bibnamefont {Paris}},
  \ and\ \bibinfo {author} {\bibfnamefont {M.~G.}\ \bibnamefont {Genoni}},\
  }\href {\doibase 10.1088/1367-2630/aa9840} {\bibfield  {journal} {\bibinfo
  {journal} {New Journal of Physics}\ }\textbf {\bibinfo {volume} {19}},\
  \bibinfo {pages} {123011} (\bibinfo {year} {2017})}\BibitemShut {NoStop}%
\bibitem [{\citenamefont {Giovannetti}\ \emph {et~al.}(2001)\citenamefont
  {Giovannetti}, \citenamefont {Lloyd},\ and\ \citenamefont
  {Maccone}}]{Giovannetti-2001}%
  \BibitemOpen
  \bibfield  {author} {\bibinfo {author} {\bibfnamefont {V.}~\bibnamefont
  {Giovannetti}}, \bibinfo {author} {\bibfnamefont {S.}~\bibnamefont {Lloyd}},
  \ and\ \bibinfo {author} {\bibfnamefont {L.}~\bibnamefont {Maccone}},\ }\href
  {\doibase 10.1038/35086525} {\bibfield  {journal} {\bibinfo  {journal}
  {Nature}\ }\textbf {\bibinfo {volume} {412}},\ \bibinfo {pages} {417}
  (\bibinfo {year} {2001})}\BibitemShut {NoStop}%
\bibitem [{\citenamefont {Boto}\ \emph {et~al.}(2000)\citenamefont {Boto},
  \citenamefont {Kok}, \citenamefont {Abrams}, \citenamefont {Braunstein},
  \citenamefont {Williams},\ and\ \citenamefont {Dowling}}]{Boto-2000}%
  \BibitemOpen
  \bibfield  {author} {\bibinfo {author} {\bibfnamefont {A.~N.}\ \bibnamefont
  {Boto}}, \bibinfo {author} {\bibfnamefont {P.}~\bibnamefont {Kok}}, \bibinfo
  {author} {\bibfnamefont {D.~S.}\ \bibnamefont {Abrams}}, \bibinfo {author}
  {\bibfnamefont {S.~L.}\ \bibnamefont {Braunstein}}, \bibinfo {author}
  {\bibfnamefont {C.~P.}\ \bibnamefont {Williams}}, \ and\ \bibinfo {author}
  {\bibfnamefont {J.~P.}\ \bibnamefont {Dowling}},\ }\href {\doibase
  10.1103/PhysRevLett.85.2733} {\bibfield  {journal} {\bibinfo  {journal}
  {Phys. Rev. Lett.}\ }\textbf {\bibinfo {volume} {85}},\ \bibinfo {pages}
  {2733} (\bibinfo {year} {2000})}\BibitemShut {NoStop}%
\bibitem [{\citenamefont {Leibfried}\ \emph {et~al.}(2004)\citenamefont
  {Leibfried}, \citenamefont {Barrett}, \citenamefont {Schaetz}, \citenamefont
  {Britton}, \citenamefont {Chiaverini}, \citenamefont {Itano}, \citenamefont
  {Jost}, \citenamefont {Langer},\ and\ \citenamefont
  {Wineland}}]{Leibfried-2004}%
  \BibitemOpen
  \bibfield  {author} {\bibinfo {author} {\bibfnamefont {D.}~\bibnamefont
  {Leibfried}}, \bibinfo {author} {\bibfnamefont {M.~D.}\ \bibnamefont
  {Barrett}}, \bibinfo {author} {\bibfnamefont {T.}~\bibnamefont {Schaetz}},
  \bibinfo {author} {\bibfnamefont {J.}~\bibnamefont {Britton}}, \bibinfo
  {author} {\bibfnamefont {J.}~\bibnamefont {Chiaverini}}, \bibinfo {author}
  {\bibfnamefont {W.~M.}\ \bibnamefont {Itano}}, \bibinfo {author}
  {\bibfnamefont {J.~D.}\ \bibnamefont {Jost}}, \bibinfo {author}
  {\bibfnamefont {C.}~\bibnamefont {Langer}}, \ and\ \bibinfo {author}
  {\bibfnamefont {D.~J.}\ \bibnamefont {Wineland}},\ }\href {\doibase
  10.1126/science.1097576} {\bibfield  {journal} {\bibinfo  {journal}
  {Science}\ }\textbf {\bibinfo {volume} {304}},\ \bibinfo {pages} {1476}
  (\bibinfo {year} {2004})}\BibitemShut {NoStop}%
\bibitem [{\citenamefont {Giovannetti}\ \emph {et~al.}(2006)\citenamefont
  {Giovannetti}, \citenamefont {Lloyd},\ and\ \citenamefont
  {Maccone}}]{Giovannetti-2006}%
  \BibitemOpen
  \bibfield  {author} {\bibinfo {author} {\bibfnamefont {V.}~\bibnamefont
  {Giovannetti}}, \bibinfo {author} {\bibfnamefont {S.}~\bibnamefont {Lloyd}},
  \ and\ \bibinfo {author} {\bibfnamefont {L.}~\bibnamefont {Maccone}},\ }\href
  {\doibase 10.1103/PhysRevLett.96.010401} {\bibfield  {journal} {\bibinfo
  {journal} {Phys. Rev. Lett.}\ }\textbf {\bibinfo {volume} {96}},\ \bibinfo
  {pages} {010401} (\bibinfo {year} {2006})}\BibitemShut {NoStop}%
\bibitem [{\citenamefont {Giovannetti}\ \emph {et~al.}(2004)\citenamefont
  {Giovannetti}, \citenamefont {Lloyd},\ and\ \citenamefont
  {Maccone}}]{Giovannetti-2004}%
  \BibitemOpen
  \bibfield  {author} {\bibinfo {author} {\bibfnamefont {V.}~\bibnamefont
  {Giovannetti}}, \bibinfo {author} {\bibfnamefont {S.}~\bibnamefont {Lloyd}},
  \ and\ \bibinfo {author} {\bibfnamefont {L.}~\bibnamefont {Maccone}},\ }\href
  {\doibase 10.1126/science.1104149} {\bibfield  {journal} {\bibinfo  {journal}
  {Science}\ }\textbf {\bibinfo {volume} {306}},\ \bibinfo {pages} {1330}
  (\bibinfo {year} {2004})}\BibitemShut {NoStop}%
\bibitem [{\citenamefont {Belliardo}\ and\ \citenamefont
  {Giovannetti}(2020)}]{Belliardo-2020}%
  \BibitemOpen
  \bibfield  {author} {\bibinfo {author} {\bibfnamefont {F.}~\bibnamefont
  {Belliardo}}\ and\ \bibinfo {author} {\bibfnamefont {V.}~\bibnamefont
  {Giovannetti}},\ }\href {\doibase 10.1103/PhysRevA.102.042613} {\bibfield
  {journal} {\bibinfo  {journal} {Phys. Rev. A}\ }\textbf {\bibinfo {volume}
  {102}},\ \bibinfo {pages} {042613} (\bibinfo {year} {2020})}\BibitemShut
  {NoStop}%
\bibitem [{\citenamefont {Roy}\ and\ \citenamefont
  {Braunstein}(2008)}]{Roy-2008}%
  \BibitemOpen
  \bibfield  {author} {\bibinfo {author} {\bibfnamefont {S.~M.}\ \bibnamefont
  {Roy}}\ and\ \bibinfo {author} {\bibfnamefont {S.~L.}\ \bibnamefont
  {Braunstein}},\ }\href {\doibase 10.1103/PhysRevLett.100.220501} {\bibfield
  {journal} {\bibinfo  {journal} {Phys. Rev. Lett.}\ }\textbf {\bibinfo
  {volume} {100}},\ \bibinfo {pages} {220501} (\bibinfo {year}
  {2008})}\BibitemShut {NoStop}%
\bibitem [{\citenamefont {Boixo}\ \emph {et~al.}(2007)\citenamefont {Boixo},
  \citenamefont {Flammia}, \citenamefont {Caves},\ and\ \citenamefont
  {Geremia}}]{boixo2007generalized}%
  \BibitemOpen
  \bibfield  {author} {\bibinfo {author} {\bibfnamefont {S.}~\bibnamefont
  {Boixo}}, \bibinfo {author} {\bibfnamefont {S.~T.}\ \bibnamefont {Flammia}},
  \bibinfo {author} {\bibfnamefont {C.~M.}\ \bibnamefont {Caves}}, \ and\
  \bibinfo {author} {\bibfnamefont {J.}~\bibnamefont {Geremia}},\ }\href
  {\doibase 10.1103/PhysRevLett.98.090401} {\bibfield  {journal} {\bibinfo
  {journal} {Phys. Rev. Lett.}\ }\textbf {\bibinfo {volume} {98}},\ \bibinfo
  {pages} {090401} (\bibinfo {year} {2007})}\BibitemShut {NoStop}%
\bibitem [{\citenamefont {De~Pasquale}\ \emph {et~al.}(2013)\citenamefont
  {De~Pasquale}, \citenamefont {Rossini}, \citenamefont {Facchi},\ and\
  \citenamefont {Giovannetti}}]{de2013quantum}%
  \BibitemOpen
  \bibfield  {author} {\bibinfo {author} {\bibfnamefont {A.}~\bibnamefont
  {De~Pasquale}}, \bibinfo {author} {\bibfnamefont {D.}~\bibnamefont
  {Rossini}}, \bibinfo {author} {\bibfnamefont {P.}~\bibnamefont {Facchi}}, \
  and\ \bibinfo {author} {\bibfnamefont {V.}~\bibnamefont {Giovannetti}},\
  }\href {\doibase 10.1103/PhysRevA.88.052117} {\bibfield  {journal} {\bibinfo
  {journal} {Phys. Rev. A}\ }\textbf {\bibinfo {volume} {88}},\ \bibinfo
  {pages} {052117} (\bibinfo {year} {2013})}\BibitemShut {NoStop}%
\bibitem [{\citenamefont {Skotiniotis}\ \emph {et~al.}(2015)\citenamefont
  {Skotiniotis}, \citenamefont {Sekatski},\ and\ \citenamefont
  {Dür}}]{skotiniotis2015quantum}%
  \BibitemOpen
  \bibfield  {author} {\bibinfo {author} {\bibfnamefont {M.}~\bibnamefont
  {Skotiniotis}}, \bibinfo {author} {\bibfnamefont {P.}~\bibnamefont
  {Sekatski}}, \ and\ \bibinfo {author} {\bibfnamefont {W.}~\bibnamefont
  {Dür}},\ }\href {\doibase 10.1088/1367-2630/17/7/073032} {\bibfield
  {journal} {\bibinfo  {journal} {New Journal of Physics}\ }\textbf {\bibinfo
  {volume} {17}},\ \bibinfo {pages} {073032} (\bibinfo {year}
  {2015})}\BibitemShut {NoStop}%
\bibitem [{\citenamefont {Pang}\ and\ \citenamefont
  {Brun}(2014)}]{pang2014quantum}%
  \BibitemOpen
  \bibfield  {author} {\bibinfo {author} {\bibfnamefont {S.}~\bibnamefont
  {Pang}}\ and\ \bibinfo {author} {\bibfnamefont {T.~A.}\ \bibnamefont
  {Brun}},\ }\href {\doibase 10.1103/PhysRevA.90.022117} {\bibfield  {journal}
  {\bibinfo  {journal} {Phys. Rev. A}\ }\textbf {\bibinfo {volume} {90}},\
  \bibinfo {pages} {022117} (\bibinfo {year} {2014})}\BibitemShut {NoStop}%
\bibitem [{\citenamefont {Albarelli}\ \emph {et~al.}(2018)\citenamefont
  {Albarelli}, \citenamefont {Rossi}, \citenamefont {Tamascelli},\ and\
  \citenamefont {Genoni}}]{Albarelli-2018}%
  \BibitemOpen
  \bibfield  {author} {\bibinfo {author} {\bibfnamefont {F.}~\bibnamefont
  {Albarelli}}, \bibinfo {author} {\bibfnamefont {M.~A.~C.}\ \bibnamefont
  {Rossi}}, \bibinfo {author} {\bibfnamefont {D.}~\bibnamefont {Tamascelli}}, \
  and\ \bibinfo {author} {\bibfnamefont {M.~G.}\ \bibnamefont {Genoni}},\
  }\href {\doibase 10.22331/q-2018-12-03-110} {\bibfield  {journal} {\bibinfo
  {journal} {{Quantum}}\ }\textbf {\bibinfo {volume} {2}},\ \bibinfo {pages}
  {110} (\bibinfo {year} {2018})}\BibitemShut {NoStop}%
\bibitem [{\citenamefont {D\"ur}\ and\ \citenamefont
  {Briegel}(2004)}]{Dur-2004}%
  \BibitemOpen
  \bibfield  {author} {\bibinfo {author} {\bibfnamefont {W.}~\bibnamefont
  {D\"ur}}\ and\ \bibinfo {author} {\bibfnamefont {H.-J.}\ \bibnamefont
  {Briegel}},\ }\href {\doibase 10.1103/PhysRevLett.92.180403} {\bibfield
  {journal} {\bibinfo  {journal} {Phys. Rev. Lett.}\ }\textbf {\bibinfo
  {volume} {92}},\ \bibinfo {pages} {180403} (\bibinfo {year}
  {2004})}\BibitemShut {NoStop}%
\bibitem [{\citenamefont {Demkowicz-Dobrza{\'{n}}ski}\ \emph
  {et~al.}(2012)\citenamefont {Demkowicz-Dobrza{\'{n}}ski}, \citenamefont
  {Ko{\l}ody{\'{n}}ski},\ and\ \citenamefont
  {Gu{\c{T}}{\u{a}}}}]{Demkowicz-2012}%
  \BibitemOpen
  \bibfield  {author} {\bibinfo {author} {\bibfnamefont {R.}~\bibnamefont
  {Demkowicz-Dobrza{\'{n}}ski}}, \bibinfo {author} {\bibfnamefont
  {J.}~\bibnamefont {Ko{\l}ody{\'{n}}ski}}, \ and\ \bibinfo {author}
  {\bibfnamefont {M.}~\bibnamefont {Gu{\c{T}}{\u{a}}}},\ }\href {\doibase
  10.1038/ncomms2067} {\bibfield  {journal} {\bibinfo  {journal} {Nature
  Communications}\ }\textbf {\bibinfo {volume} {3}},\ \bibinfo {pages} {1063}
  (\bibinfo {year} {2012})}\BibitemShut {NoStop}%
\bibitem [{\citenamefont {Matsuzaki}\ \emph {et~al.}(2011)\citenamefont
  {Matsuzaki}, \citenamefont {Benjamin},\ and\ \citenamefont
  {Fitzsimons}}]{Matsuzaki-2011}%
  \BibitemOpen
  \bibfield  {author} {\bibinfo {author} {\bibfnamefont {Y.}~\bibnamefont
  {Matsuzaki}}, \bibinfo {author} {\bibfnamefont {S.~C.}\ \bibnamefont
  {Benjamin}}, \ and\ \bibinfo {author} {\bibfnamefont {J.}~\bibnamefont
  {Fitzsimons}},\ }\href {\doibase 10.1103/PhysRevA.84.012103} {\bibfield
  {journal} {\bibinfo  {journal} {Phys. Rev. A}\ }\textbf {\bibinfo {volume}
  {84}},\ \bibinfo {pages} {012103} (\bibinfo {year} {2011})}\BibitemShut
  {NoStop}%
\bibitem [{\citenamefont {Shaji}\ and\ \citenamefont
  {Caves}(2007)}]{Shaji-2007}%
  \BibitemOpen
  \bibfield  {author} {\bibinfo {author} {\bibfnamefont {A.}~\bibnamefont
  {Shaji}}\ and\ \bibinfo {author} {\bibfnamefont {C.~M.}\ \bibnamefont
  {Caves}},\ }\href {\doibase 10.1103/PhysRevA.76.032111} {\bibfield  {journal}
  {\bibinfo  {journal} {Phys. Rev. A}\ }\textbf {\bibinfo {volume} {76}},\
  \bibinfo {pages} {032111} (\bibinfo {year} {2007})}\BibitemShut {NoStop}%
\bibitem [{\citenamefont {Huelga}\ \emph {et~al.}(1997)\citenamefont {Huelga},
  \citenamefont {Macchiavello}, \citenamefont {Pellizzari}, \citenamefont
  {Ekert}, \citenamefont {Plenio},\ and\ \citenamefont {Cirac}}]{Huelga-1997}%
  \BibitemOpen
  \bibfield  {author} {\bibinfo {author} {\bibfnamefont {S.~F.}\ \bibnamefont
  {Huelga}}, \bibinfo {author} {\bibfnamefont {C.}~\bibnamefont
  {Macchiavello}}, \bibinfo {author} {\bibfnamefont {T.}~\bibnamefont
  {Pellizzari}}, \bibinfo {author} {\bibfnamefont {A.~K.}\ \bibnamefont
  {Ekert}}, \bibinfo {author} {\bibfnamefont {M.~B.}\ \bibnamefont {Plenio}}, \
  and\ \bibinfo {author} {\bibfnamefont {J.~I.}\ \bibnamefont {Cirac}},\ }\href
  {\doibase 10.1103/PhysRevLett.79.3865} {\bibfield  {journal} {\bibinfo
  {journal} {Phys. Rev. Lett.}\ }\textbf {\bibinfo {volume} {79}},\ \bibinfo
  {pages} {3865} (\bibinfo {year} {1997})}\BibitemShut {NoStop}%
\bibitem [{\citenamefont {Cai}\ \emph {et~al.}(2012)\citenamefont {Cai},
  \citenamefont {Caruso},\ and\ \citenamefont {Plenio}}]{cai2012quantum}%
  \BibitemOpen
  \bibfield  {author} {\bibinfo {author} {\bibfnamefont {J.}~\bibnamefont
  {Cai}}, \bibinfo {author} {\bibfnamefont {F.}~\bibnamefont {Caruso}}, \ and\
  \bibinfo {author} {\bibfnamefont {M.~B.}\ \bibnamefont {Plenio}},\ }\href
  {\doibase 10.1103/PhysRevA.85.040304} {\bibfield  {journal} {\bibinfo
  {journal} {Phys. Rev. A}\ }\textbf {\bibinfo {volume} {85}},\ \bibinfo
  {pages} {040304} (\bibinfo {year} {2012})}\BibitemShut {NoStop}%
\bibitem [{\citenamefont {Liu}\ and\ \citenamefont
  {Yuan}(2017{\natexlab{a}})}]{liu2017quantum}%
  \BibitemOpen
  \bibfield  {author} {\bibinfo {author} {\bibfnamefont {J.}~\bibnamefont
  {Liu}}\ and\ \bibinfo {author} {\bibfnamefont {H.}~\bibnamefont {Yuan}},\
  }\href {\doibase 10.1103/PhysRevA.96.012117} {\bibfield  {journal} {\bibinfo
  {journal} {Phys. Rev. A}\ }\textbf {\bibinfo {volume} {96}},\ \bibinfo
  {pages} {012117} (\bibinfo {year} {2017}{\natexlab{a}})}\BibitemShut
  {NoStop}%
\bibitem [{\citenamefont {Liu}\ and\ \citenamefont
  {Yuan}(2017{\natexlab{b}})}]{liu2017control}%
  \BibitemOpen
  \bibfield  {author} {\bibinfo {author} {\bibfnamefont {J.}~\bibnamefont
  {Liu}}\ and\ \bibinfo {author} {\bibfnamefont {H.}~\bibnamefont {Yuan}},\
  }\href {\doibase 10.1103/PhysRevA.96.042114} {\bibfield  {journal} {\bibinfo
  {journal} {Phys. Rev. A}\ }\textbf {\bibinfo {volume} {96}},\ \bibinfo
  {pages} {042114} (\bibinfo {year} {2017}{\natexlab{b}})}\BibitemShut
  {NoStop}%
\bibitem [{\citenamefont {Sekatski}\ \emph {et~al.}(2017)\citenamefont
  {Sekatski}, \citenamefont {Skotiniotis}, \citenamefont
  {Ko{\l{}}ody{\'{n}}ski},\ and\ \citenamefont
  {D{\"{u}}r}}]{sekatski2017quantum}%
  \BibitemOpen
  \bibfield  {author} {\bibinfo {author} {\bibfnamefont {P.}~\bibnamefont
  {Sekatski}}, \bibinfo {author} {\bibfnamefont {M.}~\bibnamefont
  {Skotiniotis}}, \bibinfo {author} {\bibfnamefont {J.}~\bibnamefont
  {Ko{\l{}}ody{\'{n}}ski}}, \ and\ \bibinfo {author} {\bibfnamefont
  {W.}~\bibnamefont {D{\"{u}}r}},\ }\href {\doibase 10.22331/q-2017-09-06-27}
  {\bibfield  {journal} {\bibinfo  {journal} {{Quantum}}\ }\textbf {\bibinfo
  {volume} {1}},\ \bibinfo {pages} {27} (\bibinfo {year} {2017})}\BibitemShut
  {NoStop}%
\bibitem [{\citenamefont {Hentschel}\ and\ \citenamefont
  {Sanders}(2011)}]{hentschel2011efficient}%
  \BibitemOpen
  \bibfield  {author} {\bibinfo {author} {\bibfnamefont {A.}~\bibnamefont
  {Hentschel}}\ and\ \bibinfo {author} {\bibfnamefont {B.~C.}\ \bibnamefont
  {Sanders}},\ }\href {\doibase 10.1103/PhysRevLett.107.233601} {\bibfield
  {journal} {\bibinfo  {journal} {Phys. Rev. Lett.}\ }\textbf {\bibinfo
  {volume} {107}},\ \bibinfo {pages} {233601} (\bibinfo {year}
  {2011})}\BibitemShut {NoStop}%
\bibitem [{\citenamefont {Xu}\ \emph {et~al.}(2019)\citenamefont {Xu},
  \citenamefont {Li}, \citenamefont {Liu}, \citenamefont {Wang}, \citenamefont
  {Yuan},\ and\ \citenamefont {Wang}}]{xu2019generalizable}%
  \BibitemOpen
  \bibfield  {author} {\bibinfo {author} {\bibfnamefont {H.}~\bibnamefont
  {Xu}}, \bibinfo {author} {\bibfnamefont {J.}~\bibnamefont {Li}}, \bibinfo
  {author} {\bibfnamefont {L.}~\bibnamefont {Liu}}, \bibinfo {author}
  {\bibfnamefont {Y.}~\bibnamefont {Wang}}, \bibinfo {author} {\bibfnamefont
  {H.}~\bibnamefont {Yuan}}, \ and\ \bibinfo {author} {\bibfnamefont
  {X.}~\bibnamefont {Wang}},\ }\href {\doibase 10.1038/s41534-019-0198-z}
  {\bibfield  {journal} {\bibinfo  {journal} {npj Quantum Information}\
  }\textbf {\bibinfo {volume} {5}},\ \bibinfo {pages} {82} (\bibinfo {year}
  {2019})}\BibitemShut {NoStop}%
\bibitem [{\citenamefont {Cimini}\ \emph {et~al.}(2019)\citenamefont {Cimini},
  \citenamefont {Gianani}, \citenamefont {Spagnolo}, \citenamefont {Leccese},
  \citenamefont {Sciarrino},\ and\ \citenamefont
  {Barbieri}}]{cimini2019calibration}%
  \BibitemOpen
  \bibfield  {author} {\bibinfo {author} {\bibfnamefont {V.}~\bibnamefont
  {Cimini}}, \bibinfo {author} {\bibfnamefont {I.}~\bibnamefont {Gianani}},
  \bibinfo {author} {\bibfnamefont {N.}~\bibnamefont {Spagnolo}}, \bibinfo
  {author} {\bibfnamefont {F.}~\bibnamefont {Leccese}}, \bibinfo {author}
  {\bibfnamefont {F.}~\bibnamefont {Sciarrino}}, \ and\ \bibinfo {author}
  {\bibfnamefont {M.}~\bibnamefont {Barbieri}},\ }\href {\doibase
  10.1103/PhysRevLett.123.230502} {\bibfield  {journal} {\bibinfo  {journal}
  {Phys. Rev. Lett.}\ }\textbf {\bibinfo {volume} {123}},\ \bibinfo {pages}
  {230502} (\bibinfo {year} {2019})}\BibitemShut {NoStop}%
\bibitem [{\citenamefont {Meyer}\ \emph {et~al.}(2020)\citenamefont {Meyer},
  \citenamefont {Borregaard},\ and\ \citenamefont
  {Eisert}}]{meyer2020variational}%
  \BibitemOpen
  \bibfield  {author} {\bibinfo {author} {\bibfnamefont {J.~J.}\ \bibnamefont
  {Meyer}}, \bibinfo {author} {\bibfnamefont {J.}~\bibnamefont {Borregaard}}, \
  and\ \bibinfo {author} {\bibfnamefont {J.}~\bibnamefont {Eisert}},\
  }\href@noop {} {\enquote {\bibinfo {title} {A variational toolbox for quantum
  multi-parameter estimation},}\ } (\bibinfo {year} {2020}),\ \Eprint
  {http://arxiv.org/abs/2006.06303} {arXiv:2006.06303 [quant-ph]} \BibitemShut
  {NoStop}%
\bibitem [{\citenamefont {Lovett}\ \emph {et~al.}(2013)\citenamefont {Lovett},
  \citenamefont {Crosnier}, \citenamefont {Perarnau-Llobet},\ and\
  \citenamefont {Sanders}}]{lovett2013differential}%
  \BibitemOpen
  \bibfield  {author} {\bibinfo {author} {\bibfnamefont {N.~B.}\ \bibnamefont
  {Lovett}}, \bibinfo {author} {\bibfnamefont {C.}~\bibnamefont {Crosnier}},
  \bibinfo {author} {\bibfnamefont {M.}~\bibnamefont {Perarnau-Llobet}}, \ and\
  \bibinfo {author} {\bibfnamefont {B.~C.}\ \bibnamefont {Sanders}},\ }\href
  {\doibase 10.1103/PhysRevLett.110.220501} {\bibfield  {journal} {\bibinfo
  {journal} {Phys. Rev. Lett.}\ }\textbf {\bibinfo {volume} {110}},\ \bibinfo
  {pages} {220501} (\bibinfo {year} {2013})}\BibitemShut {NoStop}%
\bibitem [{\citenamefont {Yuan}\ and\ \citenamefont
  {Fung}(2015)}]{yuan2015optimal}%
  \BibitemOpen
  \bibfield  {author} {\bibinfo {author} {\bibfnamefont {H.}~\bibnamefont
  {Yuan}}\ and\ \bibinfo {author} {\bibfnamefont {C.-H.~F.}\ \bibnamefont
  {Fung}},\ }\href {\doibase 10.1103/PhysRevLett.115.110401} {\bibfield
  {journal} {\bibinfo  {journal} {Phys. Rev. Lett.}\ }\textbf {\bibinfo
  {volume} {115}},\ \bibinfo {pages} {110401} (\bibinfo {year}
  {2015})}\BibitemShut {NoStop}%
\bibitem [{\citenamefont {Pang}\ and\ \citenamefont
  {Jordan}(2017)}]{pang2017optimal}%
  \BibitemOpen
  \bibfield  {author} {\bibinfo {author} {\bibfnamefont {S.}~\bibnamefont
  {Pang}}\ and\ \bibinfo {author} {\bibfnamefont {A.~N.}\ \bibnamefont
  {Jordan}},\ }\href {\doibase 10.1038/ncomms14695} {\bibfield  {journal}
  {\bibinfo  {journal} {Nature Communications}\ }\textbf {\bibinfo {volume}
  {8}},\ \bibinfo {pages} {14695} (\bibinfo {year} {2017})}\BibitemShut
  {NoStop}%
\bibitem [{\citenamefont {Naghiloo}\ \emph {et~al.}(2017)\citenamefont
  {Naghiloo}, \citenamefont {Jordan},\ and\ \citenamefont
  {Murch}}]{naghiloo2017achieving}%
  \BibitemOpen
  \bibfield  {author} {\bibinfo {author} {\bibfnamefont {M.}~\bibnamefont
  {Naghiloo}}, \bibinfo {author} {\bibfnamefont {A.~N.}\ \bibnamefont
  {Jordan}}, \ and\ \bibinfo {author} {\bibfnamefont {K.~W.}\ \bibnamefont
  {Murch}},\ }\href {\doibase 10.1103/PhysRevLett.119.180801} {\bibfield
  {journal} {\bibinfo  {journal} {Phys. Rev. Lett.}\ }\textbf {\bibinfo
  {volume} {119}},\ \bibinfo {pages} {180801} (\bibinfo {year}
  {2017})}\BibitemShut {NoStop}%
\bibitem [{\citenamefont {Fiderer}\ and\ \citenamefont
  {Braun}(2018)}]{fiderer2018quantum}%
  \BibitemOpen
  \bibfield  {author} {\bibinfo {author} {\bibfnamefont {L.~J.}\ \bibnamefont
  {Fiderer}}\ and\ \bibinfo {author} {\bibfnamefont {D.}~\bibnamefont
  {Braun}},\ }\href {\doibase 10.1038/s41467-018-03623-z} {\bibfield  {journal}
  {\bibinfo  {journal} {Nature Communications}\ }\textbf {\bibinfo {volume}
  {9}},\ \bibinfo {pages} {1351} (\bibinfo {year} {2018})}\BibitemShut
  {NoStop}%
\bibitem [{\citenamefont {Lang}\ \emph {et~al.}(2015)\citenamefont {Lang},
  \citenamefont {Liu},\ and\ \citenamefont {Monteiro}}]{lang2015dynamical}%
  \BibitemOpen
  \bibfield  {author} {\bibinfo {author} {\bibfnamefont {J.~E.}\ \bibnamefont
  {Lang}}, \bibinfo {author} {\bibfnamefont {R.~B.}\ \bibnamefont {Liu}}, \
  and\ \bibinfo {author} {\bibfnamefont {T.~S.}\ \bibnamefont {Monteiro}},\
  }\href {\doibase 10.1103/PhysRevX.5.041016} {\bibfield  {journal} {\bibinfo
  {journal} {Phys. Rev. X}\ }\textbf {\bibinfo {volume} {5}},\ \bibinfo {pages}
  {041016} (\bibinfo {year} {2015})}\BibitemShut {NoStop}%
\bibitem [{\citenamefont {Mishra}\ and\ \citenamefont
  {Bayat}(2020)}]{mishra2020driving}%
  \BibitemOpen
  \bibfield  {author} {\bibinfo {author} {\bibfnamefont {U.}~\bibnamefont
  {Mishra}}\ and\ \bibinfo {author} {\bibfnamefont {A.}~\bibnamefont {Bayat}},\
  }\href@noop {} {\enquote {\bibinfo {title} {Driving enhanced quantum sensing
  in partially accessible many-body systems},}\ } (\bibinfo {year} {2020}),\
  \Eprint {http://arxiv.org/abs/2010.09050} {arXiv:2010.09050 [quant-ph]}
  \BibitemShut {NoStop}%
\bibitem [{\citenamefont {Bonato}\ \emph {et~al.}(2016)\citenamefont {Bonato},
  \citenamefont {Blok}, \citenamefont {Dinani}, \citenamefont {Berry},
  \citenamefont {Markham}, \citenamefont {Twitchen},\ and\ \citenamefont
  {Hanson}}]{bonato2016optimized}%
  \BibitemOpen
  \bibfield  {author} {\bibinfo {author} {\bibfnamefont {C.}~\bibnamefont
  {Bonato}}, \bibinfo {author} {\bibfnamefont {M.~S.}\ \bibnamefont {Blok}},
  \bibinfo {author} {\bibfnamefont {H.~T.}\ \bibnamefont {Dinani}}, \bibinfo
  {author} {\bibfnamefont {D.~W.}\ \bibnamefont {Berry}}, \bibinfo {author}
  {\bibfnamefont {M.~L.}\ \bibnamefont {Markham}}, \bibinfo {author}
  {\bibfnamefont {D.~J.}\ \bibnamefont {Twitchen}}, \ and\ \bibinfo {author}
  {\bibfnamefont {R.}~\bibnamefont {Hanson}},\ }\href {\doibase
  10.1038/nnano.2015.261} {\bibfield  {journal} {\bibinfo  {journal} {Nature
  Nanotechnology}\ }\textbf {\bibinfo {volume} {11}},\ \bibinfo {pages} {247}
  (\bibinfo {year} {2016})}\BibitemShut {NoStop}%
\bibitem [{\citenamefont {Higgins}\ \emph
  {et~al.}(2007{\natexlab{a}})\citenamefont {Higgins}, \citenamefont {Berry},
  \citenamefont {Bartlett}, \citenamefont {Wiseman},\ and\ \citenamefont
  {Pryde}}]{higgins2007entanglement}%
  \BibitemOpen
  \bibfield  {author} {\bibinfo {author} {\bibfnamefont {B.~L.}\ \bibnamefont
  {Higgins}}, \bibinfo {author} {\bibfnamefont {D.~W.}\ \bibnamefont {Berry}},
  \bibinfo {author} {\bibfnamefont {S.~D.}\ \bibnamefont {Bartlett}}, \bibinfo
  {author} {\bibfnamefont {H.~M.}\ \bibnamefont {Wiseman}}, \ and\ \bibinfo
  {author} {\bibfnamefont {G.~J.}\ \bibnamefont {Pryde}},\ }\href {\doibase
  10.1038/nature06257} {\bibfield  {journal} {\bibinfo  {journal} {Nature}\
  }\textbf {\bibinfo {volume} {450}},\ \bibinfo {pages} {393} (\bibinfo {year}
  {2007}{\natexlab{a}})}\BibitemShut {NoStop}%
\bibitem [{\citenamefont {Berry}\ \emph {et~al.}(2009)\citenamefont {Berry},
  \citenamefont {Higgins}, \citenamefont {Bartlett}, \citenamefont {Mitchell},
  \citenamefont {Pryde},\ and\ \citenamefont {Wiseman}}]{berry2009perform}%
  \BibitemOpen
  \bibfield  {author} {\bibinfo {author} {\bibfnamefont {D.~W.}\ \bibnamefont
  {Berry}}, \bibinfo {author} {\bibfnamefont {B.~L.}\ \bibnamefont {Higgins}},
  \bibinfo {author} {\bibfnamefont {S.~D.}\ \bibnamefont {Bartlett}}, \bibinfo
  {author} {\bibfnamefont {M.~W.}\ \bibnamefont {Mitchell}}, \bibinfo {author}
  {\bibfnamefont {G.~J.}\ \bibnamefont {Pryde}}, \ and\ \bibinfo {author}
  {\bibfnamefont {H.~M.}\ \bibnamefont {Wiseman}},\ }\href {\doibase
  10.1103/PhysRevA.80.052114} {\bibfield  {journal} {\bibinfo  {journal} {Phys.
  Rev. A}\ }\textbf {\bibinfo {volume} {80}},\ \bibinfo {pages} {052114}
  (\bibinfo {year} {2009})}\BibitemShut {NoStop}%
\bibitem [{\citenamefont {Jones}\ \emph {et~al.}(2020)\citenamefont {Jones},
  \citenamefont {Bose},\ and\ \citenamefont {Bayat}}]{jones2020remote}%
  \BibitemOpen
  \bibfield  {author} {\bibinfo {author} {\bibfnamefont {G.~S.}\ \bibnamefont
  {Jones}}, \bibinfo {author} {\bibfnamefont {S.}~\bibnamefont {Bose}}, \ and\
  \bibinfo {author} {\bibfnamefont {A.}~\bibnamefont {Bayat}},\ }\href@noop {}
  {\enquote {\bibinfo {title} {Remote quantum sensing with heisenberg limited
  sensitivity in many body systems},}\ } (\bibinfo {year} {2020}),\ \Eprint
  {http://arxiv.org/abs/2003.02308} {arXiv:2003.02308 [quant-ph]} \BibitemShut
  {NoStop}%
\bibitem [{\citenamefont {Beau}\ and\ \citenamefont {del
  Campo}(2017)}]{Beau-2017}%
  \BibitemOpen
  \bibfield  {author} {\bibinfo {author} {\bibfnamefont {M.}~\bibnamefont
  {Beau}}\ and\ \bibinfo {author} {\bibfnamefont {A.}~\bibnamefont {del
  Campo}},\ }\href {\doibase 10.1103/PhysRevLett.119.010403} {\bibfield
  {journal} {\bibinfo  {journal} {Phys. Rev. Lett.}\ }\textbf {\bibinfo
  {volume} {119}},\ \bibinfo {pages} {010403} (\bibinfo {year}
  {2017})}\BibitemShut {NoStop}%
\bibitem [{\citenamefont {Guo}\ \emph {et~al.}(2016)\citenamefont {Guo},
  \citenamefont {Xu}, \citenamefont {Zou},\ and\ \citenamefont
  {Shao}}]{Guo-2016}%
  \BibitemOpen
  \bibfield  {author} {\bibinfo {author} {\bibfnamefont {L.-S.}\ \bibnamefont
  {Guo}}, \bibinfo {author} {\bibfnamefont {B.-M.}\ \bibnamefont {Xu}},
  \bibinfo {author} {\bibfnamefont {J.}~\bibnamefont {Zou}}, \ and\ \bibinfo
  {author} {\bibfnamefont {B.}~\bibnamefont {Shao}},\ }\href {\doibase
  10.1038/srep33254} {\bibfield  {journal} {\bibinfo  {journal} {Scientific
  Reports}\ }\textbf {\bibinfo {volume} {6}},\ \bibinfo {pages} {33254}
  (\bibinfo {year} {2016})}\BibitemShut {NoStop}%
\bibitem [{\citenamefont {Rey}\ \emph {et~al.}(2007)\citenamefont {Rey},
  \citenamefont {Jiang},\ and\ \citenamefont {Lukin}}]{Rey-2007}%
  \BibitemOpen
  \bibfield  {author} {\bibinfo {author} {\bibfnamefont {A.~M.}\ \bibnamefont
  {Rey}}, \bibinfo {author} {\bibfnamefont {L.}~\bibnamefont {Jiang}}, \ and\
  \bibinfo {author} {\bibfnamefont {M.~D.}\ \bibnamefont {Lukin}},\ }\href
  {\doibase 10.1103/PhysRevA.76.053617} {\bibfield  {journal} {\bibinfo
  {journal} {Phys. Rev. A}\ }\textbf {\bibinfo {volume} {76}},\ \bibinfo
  {pages} {053617} (\bibinfo {year} {2007})}\BibitemShut {NoStop}%
\bibitem [{\citenamefont {Choi}\ and\ \citenamefont
  {Sundaram}(2008)}]{Choi-2008}%
  \BibitemOpen
  \bibfield  {author} {\bibinfo {author} {\bibfnamefont {S.}~\bibnamefont
  {Choi}}\ and\ \bibinfo {author} {\bibfnamefont {B.}~\bibnamefont
  {Sundaram}},\ }\href {\doibase 10.1103/PhysRevA.77.053613} {\bibfield
  {journal} {\bibinfo  {journal} {Phys. Rev. A}\ }\textbf {\bibinfo {volume}
  {77}},\ \bibinfo {pages} {053613} (\bibinfo {year} {2008})}\BibitemShut
  {NoStop}%
\bibitem [{\citenamefont {Boixo}\ \emph {et~al.}(2008)\citenamefont {Boixo},
  \citenamefont {Datta}, \citenamefont {Davis}, \citenamefont {Flammia},
  \citenamefont {Shaji},\ and\ \citenamefont {Caves}}]{Boixo-2008}%
  \BibitemOpen
  \bibfield  {author} {\bibinfo {author} {\bibfnamefont {S.}~\bibnamefont
  {Boixo}}, \bibinfo {author} {\bibfnamefont {A.}~\bibnamefont {Datta}},
  \bibinfo {author} {\bibfnamefont {M.~J.}\ \bibnamefont {Davis}}, \bibinfo
  {author} {\bibfnamefont {S.~T.}\ \bibnamefont {Flammia}}, \bibinfo {author}
  {\bibfnamefont {A.}~\bibnamefont {Shaji}}, \ and\ \bibinfo {author}
  {\bibfnamefont {C.~M.}\ \bibnamefont {Caves}},\ }\href {\doibase
  10.1103/PhysRevLett.101.040403} {\bibfield  {journal} {\bibinfo  {journal}
  {Phys. Rev. Lett.}\ }\textbf {\bibinfo {volume} {101}},\ \bibinfo {pages}
  {040403} (\bibinfo {year} {2008})}\BibitemShut {NoStop}%
\bibitem [{\citenamefont {Boixo}\ \emph {et~al.}(2009)\citenamefont {Boixo},
  \citenamefont {Datta}, \citenamefont {Davis}, \citenamefont {Shaji},
  \citenamefont {Tacla},\ and\ \citenamefont {Caves}}]{Boixo-2009}%
  \BibitemOpen
  \bibfield  {author} {\bibinfo {author} {\bibfnamefont {S.}~\bibnamefont
  {Boixo}}, \bibinfo {author} {\bibfnamefont {A.}~\bibnamefont {Datta}},
  \bibinfo {author} {\bibfnamefont {M.~J.}\ \bibnamefont {Davis}}, \bibinfo
  {author} {\bibfnamefont {A.}~\bibnamefont {Shaji}}, \bibinfo {author}
  {\bibfnamefont {A.~B.}\ \bibnamefont {Tacla}}, \ and\ \bibinfo {author}
  {\bibfnamefont {C.~M.}\ \bibnamefont {Caves}},\ }\href {\doibase
  10.1103/PhysRevA.80.032103} {\bibfield  {journal} {\bibinfo  {journal} {Phys.
  Rev. A}\ }\textbf {\bibinfo {volume} {80}},\ \bibinfo {pages} {032103}
  (\bibinfo {year} {2009})}\BibitemShut {NoStop}%
\bibitem [{\citenamefont {Czajkowski}\ \emph {et~al.}(2019)\citenamefont
  {Czajkowski}, \citenamefont {Paw{\l}owski},\ and\ \citenamefont
  {Demkowicz-Dobrza{\'{n}}ski}}]{Czajkowski-2019}%
  \BibitemOpen
  \bibfield  {author} {\bibinfo {author} {\bibfnamefont {J.}~\bibnamefont
  {Czajkowski}}, \bibinfo {author} {\bibfnamefont {K.}~\bibnamefont
  {Paw{\l}owski}}, \ and\ \bibinfo {author} {\bibfnamefont {R.}~\bibnamefont
  {Demkowicz-Dobrza{\'{n}}ski}},\ }\href {\doibase 10.1088/1367-2630/ab1fc2}
  {\bibfield  {journal} {\bibinfo  {journal} {New Journal of Physics}\ }\textbf
  {\bibinfo {volume} {21}},\ \bibinfo {pages} {053031} (\bibinfo {year}
  {2019})}\BibitemShut {NoStop}%
\bibitem [{\citenamefont {Rams}\ \emph {et~al.}(2018)\citenamefont {Rams},
  \citenamefont {Sierant}, \citenamefont {Dutta}, \citenamefont {Horodecki},\
  and\ \citenamefont {Zakrzewski}}]{Rams-2018}%
  \BibitemOpen
  \bibfield  {author} {\bibinfo {author} {\bibfnamefont {M.~M.}\ \bibnamefont
  {Rams}}, \bibinfo {author} {\bibfnamefont {P.}~\bibnamefont {Sierant}},
  \bibinfo {author} {\bibfnamefont {O.}~\bibnamefont {Dutta}}, \bibinfo
  {author} {\bibfnamefont {P.}~\bibnamefont {Horodecki}}, \ and\ \bibinfo
  {author} {\bibfnamefont {J.}~\bibnamefont {Zakrzewski}},\ }\href {\doibase
  10.1103/PhysRevX.8.021022} {\bibfield  {journal} {\bibinfo  {journal} {Phys.
  Rev. X}\ }\textbf {\bibinfo {volume} {8}},\ \bibinfo {pages} {021022}
  (\bibinfo {year} {2018})}\BibitemShut {NoStop}%
\bibitem [{\citenamefont {Invernizzi}\ \emph {et~al.}(2008)\citenamefont
  {Invernizzi}, \citenamefont {Korbman}, \citenamefont {Campos~Venuti},\ and\
  \citenamefont {Paris}}]{Invernizzi-2008}%
  \BibitemOpen
  \bibfield  {author} {\bibinfo {author} {\bibfnamefont {C.}~\bibnamefont
  {Invernizzi}}, \bibinfo {author} {\bibfnamefont {M.}~\bibnamefont {Korbman}},
  \bibinfo {author} {\bibfnamefont {L.}~\bibnamefont {Campos~Venuti}}, \ and\
  \bibinfo {author} {\bibfnamefont {M.~G.~A.}\ \bibnamefont {Paris}},\ }\href
  {\doibase 10.1103/PhysRevA.78.042106} {\bibfield  {journal} {\bibinfo
  {journal} {Phys. Rev. A}\ }\textbf {\bibinfo {volume} {78}},\ \bibinfo
  {pages} {042106} (\bibinfo {year} {2008})}\BibitemShut {NoStop}%
\bibitem [{\citenamefont {Zanardi}\ \emph {et~al.}(2008)\citenamefont
  {Zanardi}, \citenamefont {Paris},\ and\ \citenamefont
  {Campos~Venuti}}]{Zanardi-2008}%
  \BibitemOpen
  \bibfield  {author} {\bibinfo {author} {\bibfnamefont {P.}~\bibnamefont
  {Zanardi}}, \bibinfo {author} {\bibfnamefont {M.~G.~A.}\ \bibnamefont
  {Paris}}, \ and\ \bibinfo {author} {\bibfnamefont {L.}~\bibnamefont
  {Campos~Venuti}},\ }\href {\doibase 10.1103/PhysRevA.78.042105} {\bibfield
  {journal} {\bibinfo  {journal} {Phys. Rev. A}\ }\textbf {\bibinfo {volume}
  {78}},\ \bibinfo {pages} {042105} (\bibinfo {year} {2008})}\BibitemShut
  {NoStop}%
\bibitem [{\citenamefont {Gammelmark}\ and\ \citenamefont
  {M{\o}lmer}(2011)}]{Gammelmark-2011}%
  \BibitemOpen
  \bibfield  {author} {\bibinfo {author} {\bibfnamefont {S.}~\bibnamefont
  {Gammelmark}}\ and\ \bibinfo {author} {\bibfnamefont {K.}~\bibnamefont
  {M{\o}lmer}},\ }\href {\doibase 10.1088/1367-2630/13/5/053035} {\bibfield
  {journal} {\bibinfo  {journal} {New Journal of Physics}\ }\textbf {\bibinfo
  {volume} {13}},\ \bibinfo {pages} {053035} (\bibinfo {year}
  {2011})}\BibitemShut {NoStop}%
\bibitem [{\citenamefont {Salvatori}\ \emph {et~al.}(2014)\citenamefont
  {Salvatori}, \citenamefont {Mandarino},\ and\ \citenamefont
  {Paris}}]{Giulio-2014}%
  \BibitemOpen
  \bibfield  {author} {\bibinfo {author} {\bibfnamefont {G.}~\bibnamefont
  {Salvatori}}, \bibinfo {author} {\bibfnamefont {A.}~\bibnamefont
  {Mandarino}}, \ and\ \bibinfo {author} {\bibfnamefont {M.~G.~A.}\
  \bibnamefont {Paris}},\ }\href {\doibase 10.1103/PhysRevA.90.022111}
  {\bibfield  {journal} {\bibinfo  {journal} {Phys. Rev. A}\ }\textbf {\bibinfo
  {volume} {90}},\ \bibinfo {pages} {022111} (\bibinfo {year}
  {2014})}\BibitemShut {NoStop}%
\bibitem [{\citenamefont {Bina}\ \emph {et~al.}(2016)\citenamefont {Bina},
  \citenamefont {Amelio},\ and\ \citenamefont {Paris}}]{Bina-2016}%
  \BibitemOpen
  \bibfield  {author} {\bibinfo {author} {\bibfnamefont {M.}~\bibnamefont
  {Bina}}, \bibinfo {author} {\bibfnamefont {I.}~\bibnamefont {Amelio}}, \ and\
  \bibinfo {author} {\bibfnamefont {M.~G.~A.}\ \bibnamefont {Paris}},\ }\href
  {\doibase 10.1103/PhysRevE.93.052118} {\bibfield  {journal} {\bibinfo
  {journal} {Phys. Rev. E}\ }\textbf {\bibinfo {volume} {93}},\ \bibinfo
  {pages} {052118} (\bibinfo {year} {2016})}\BibitemShut {NoStop}%
\bibitem [{\citenamefont {Fr\'erot}\ and\ \citenamefont
  {Roscilde}(2018)}]{Frerot-2018}%
  \BibitemOpen
  \bibfield  {author} {\bibinfo {author} {\bibfnamefont {I.}~\bibnamefont
  {Fr\'erot}}\ and\ \bibinfo {author} {\bibfnamefont {T.}~\bibnamefont
  {Roscilde}},\ }\href {\doibase 10.1103/PhysRevLett.121.020402} {\bibfield
  {journal} {\bibinfo  {journal} {Phys. Rev. Lett.}\ }\textbf {\bibinfo
  {volume} {121}},\ \bibinfo {pages} {020402} (\bibinfo {year}
  {2018})}\BibitemShut {NoStop}%
\bibitem [{\citenamefont {Chu}\ \emph {et~al.}(2021)\citenamefont {Chu},
  \citenamefont {Zhang}, \citenamefont {Yu},\ and\ \citenamefont
  {Cai}}]{Chu-2021}%
  \BibitemOpen
  \bibfield  {author} {\bibinfo {author} {\bibfnamefont {Y.}~\bibnamefont
  {Chu}}, \bibinfo {author} {\bibfnamefont {S.}~\bibnamefont {Zhang}}, \bibinfo
  {author} {\bibfnamefont {B.}~\bibnamefont {Yu}}, \ and\ \bibinfo {author}
  {\bibfnamefont {J.}~\bibnamefont {Cai}},\ }\href {\doibase
  10.1103/PhysRevLett.126.010502} {\bibfield  {journal} {\bibinfo  {journal}
  {Phys. Rev. Lett.}\ }\textbf {\bibinfo {volume} {126}},\ \bibinfo {pages}
  {010502} (\bibinfo {year} {2021})}\BibitemShut {NoStop}%
\bibitem [{\citenamefont {Garbe}\ \emph {et~al.}(2020)\citenamefont {Garbe},
  \citenamefont {Bina}, \citenamefont {Keller}, \citenamefont {Paris},\ and\
  \citenamefont {Felicetti}}]{Garbe-2020}%
  \BibitemOpen
  \bibfield  {author} {\bibinfo {author} {\bibfnamefont {L.}~\bibnamefont
  {Garbe}}, \bibinfo {author} {\bibfnamefont {M.}~\bibnamefont {Bina}},
  \bibinfo {author} {\bibfnamefont {A.}~\bibnamefont {Keller}}, \bibinfo
  {author} {\bibfnamefont {M.~G.~A.}\ \bibnamefont {Paris}}, \ and\ \bibinfo
  {author} {\bibfnamefont {S.}~\bibnamefont {Felicetti}},\ }\href {\doibase
  10.1103/PhysRevLett.124.120504} {\bibfield  {journal} {\bibinfo  {journal}
  {Phys. Rev. Lett.}\ }\textbf {\bibinfo {volume} {124}},\ \bibinfo {pages}
  {120504} (\bibinfo {year} {2020})}\BibitemShut {NoStop}%
\bibitem [{\citenamefont {Rossi}\ \emph {et~al.}(2017)\citenamefont {Rossi},
  \citenamefont {Bina}, \citenamefont {Paris}, \citenamefont {Genoni},
  \citenamefont {Adesso},\ and\ \citenamefont {Tufarelli}}]{Rossi-2017}%
  \BibitemOpen
  \bibfield  {author} {\bibinfo {author} {\bibfnamefont {M.~A.~C.}\
  \bibnamefont {Rossi}}, \bibinfo {author} {\bibfnamefont {M.}~\bibnamefont
  {Bina}}, \bibinfo {author} {\bibfnamefont {M.~G.~A.}\ \bibnamefont {Paris}},
  \bibinfo {author} {\bibfnamefont {M.~G.}\ \bibnamefont {Genoni}}, \bibinfo
  {author} {\bibfnamefont {G.}~\bibnamefont {Adesso}}, \ and\ \bibinfo {author}
  {\bibfnamefont {T.}~\bibnamefont {Tufarelli}},\ }\href {\doibase
  10.1088/2058-9565/aa540a} {\bibfield  {journal} {\bibinfo  {journal} {Quantum
  Science and Technology}\ }\textbf {\bibinfo {volume} {2}},\ \bibinfo {pages}
  {01LT01} (\bibinfo {year} {2017})}\BibitemShut {NoStop}%
\bibitem [{\citenamefont {Mehboudi}\ \emph {et~al.}(2016)\citenamefont
  {Mehboudi}, \citenamefont {Correa},\ and\ \citenamefont
  {Sanpera}}]{Mehboudi-2016}%
  \BibitemOpen
  \bibfield  {author} {\bibinfo {author} {\bibfnamefont {M.}~\bibnamefont
  {Mehboudi}}, \bibinfo {author} {\bibfnamefont {L.~A.}\ \bibnamefont
  {Correa}}, \ and\ \bibinfo {author} {\bibfnamefont {A.}~\bibnamefont
  {Sanpera}},\ }\href {\doibase 10.1103/PhysRevA.94.042121} {\bibfield
  {journal} {\bibinfo  {journal} {Phys. Rev. A}\ }\textbf {\bibinfo {volume}
  {94}},\ \bibinfo {pages} {042121} (\bibinfo {year} {2016})}\BibitemShut
  {NoStop}%
\bibitem [{\citenamefont {Okamoto}\ \emph {et~al.}(2012)\citenamefont
  {Okamoto}, \citenamefont {Iefuji}, \citenamefont {Oyama}, \citenamefont
  {Yamagata}, \citenamefont {Imai}, \citenamefont {Fujiwara},\ and\
  \citenamefont {Takeuchi}}]{Okamoto-2012}%
  \BibitemOpen
  \bibfield  {author} {\bibinfo {author} {\bibfnamefont {R.}~\bibnamefont
  {Okamoto}}, \bibinfo {author} {\bibfnamefont {M.}~\bibnamefont {Iefuji}},
  \bibinfo {author} {\bibfnamefont {S.}~\bibnamefont {Oyama}}, \bibinfo
  {author} {\bibfnamefont {K.}~\bibnamefont {Yamagata}}, \bibinfo {author}
  {\bibfnamefont {H.}~\bibnamefont {Imai}}, \bibinfo {author} {\bibfnamefont
  {A.}~\bibnamefont {Fujiwara}}, \ and\ \bibinfo {author} {\bibfnamefont
  {S.}~\bibnamefont {Takeuchi}},\ }\href {\doibase
  10.1103/PhysRevLett.109.130404} {\bibfield  {journal} {\bibinfo  {journal}
  {Phys. Rev. Lett.}\ }\textbf {\bibinfo {volume} {109}},\ \bibinfo {pages}
  {130404} (\bibinfo {year} {2012})}\BibitemShut {NoStop}%
\bibitem [{\citenamefont {Okamoto}\ \emph {et~al.}(2017)\citenamefont
  {Okamoto}, \citenamefont {Oyama}, \citenamefont {Yamagata}, \citenamefont
  {Fujiwara},\ and\ \citenamefont {Takeuchi}}]{Okamoto-2017}%
  \BibitemOpen
  \bibfield  {author} {\bibinfo {author} {\bibfnamefont {R.}~\bibnamefont
  {Okamoto}}, \bibinfo {author} {\bibfnamefont {S.}~\bibnamefont {Oyama}},
  \bibinfo {author} {\bibfnamefont {K.}~\bibnamefont {Yamagata}}, \bibinfo
  {author} {\bibfnamefont {A.}~\bibnamefont {Fujiwara}}, \ and\ \bibinfo
  {author} {\bibfnamefont {S.}~\bibnamefont {Takeuchi}},\ }\href {\doibase
  10.1103/PhysRevA.96.022124} {\bibfield  {journal} {\bibinfo  {journal} {Phys.
  Rev. A}\ }\textbf {\bibinfo {volume} {96}},\ \bibinfo {pages} {022124}
  (\bibinfo {year} {2017})}\BibitemShut {NoStop}%
\bibitem [{\citenamefont {Fujiwara}(2006)}]{Fujiwara-2006}%
  \BibitemOpen
  \bibfield  {author} {\bibinfo {author} {\bibfnamefont {A.}~\bibnamefont
  {Fujiwara}},\ }\href {\doibase 10.1088/0305-4470/39/40/014} {\bibfield
  {journal} {\bibinfo  {journal} {Journal of Physics A: Mathematical and
  General}\ }\textbf {\bibinfo {volume} {39}},\ \bibinfo {pages} {12489}
  (\bibinfo {year} {2006})}\BibitemShut {NoStop}%
\bibitem [{\citenamefont {Fujiwara}(2011)}]{Fujiwara-2011}%
  \BibitemOpen
  \bibfield  {author} {\bibinfo {author} {\bibfnamefont {A.}~\bibnamefont
  {Fujiwara}},\ }\href {\doibase 10.1088/1751-8113/44/7/079501} {\bibfield
  {journal} {\bibinfo  {journal} {Journal of Physics A: Mathematical and
  Theoretical}\ }\textbf {\bibinfo {volume} {44}},\ \bibinfo {pages} {079501}
  (\bibinfo {year} {2011})}\BibitemShut {NoStop}%
\bibitem [{\citenamefont {Wiseman}(1995)}]{Wiseman-1995}%
  \BibitemOpen
  \bibfield  {author} {\bibinfo {author} {\bibfnamefont {H.~M.}\ \bibnamefont
  {Wiseman}},\ }\href {\doibase 10.1103/PhysRevLett.75.4587} {\bibfield
  {journal} {\bibinfo  {journal} {Phys. Rev. Lett.}\ }\textbf {\bibinfo
  {volume} {75}},\ \bibinfo {pages} {4587} (\bibinfo {year}
  {1995})}\BibitemShut {NoStop}%
\bibitem [{\citenamefont {Higgins}\ \emph
  {et~al.}(2007{\natexlab{b}})\citenamefont {Higgins}, \citenamefont {Berry},
  \citenamefont {Bartlett}, \citenamefont {Wiseman},\ and\ \citenamefont
  {Pryde}}]{Higgins-2007}%
  \BibitemOpen
  \bibfield  {author} {\bibinfo {author} {\bibfnamefont {B.~L.}\ \bibnamefont
  {Higgins}}, \bibinfo {author} {\bibfnamefont {D.~W.}\ \bibnamefont {Berry}},
  \bibinfo {author} {\bibfnamefont {S.~D.}\ \bibnamefont {Bartlett}}, \bibinfo
  {author} {\bibfnamefont {H.~M.}\ \bibnamefont {Wiseman}}, \ and\ \bibinfo
  {author} {\bibfnamefont {G.~J.}\ \bibnamefont {Pryde}},\ }\href {\doibase
  10.1038/nature06257} {\bibfield  {journal} {\bibinfo  {journal} {Nature}\
  }\textbf {\bibinfo {volume} {450}},\ \bibinfo {pages} {393} (\bibinfo {year}
  {2007}{\natexlab{b}})}\BibitemShut {NoStop}%
\bibitem [{\citenamefont {Armen}\ \emph {et~al.}(2002)\citenamefont {Armen},
  \citenamefont {Au}, \citenamefont {Stockton}, \citenamefont {Doherty},\ and\
  \citenamefont {Mabuchi}}]{Armen-2002}%
  \BibitemOpen
  \bibfield  {author} {\bibinfo {author} {\bibfnamefont {M.~A.}\ \bibnamefont
  {Armen}}, \bibinfo {author} {\bibfnamefont {J.~K.}\ \bibnamefont {Au}},
  \bibinfo {author} {\bibfnamefont {J.~K.}\ \bibnamefont {Stockton}}, \bibinfo
  {author} {\bibfnamefont {A.~C.}\ \bibnamefont {Doherty}}, \ and\ \bibinfo
  {author} {\bibfnamefont {H.}~\bibnamefont {Mabuchi}},\ }\href {\doibase
  10.1103/PhysRevLett.89.133602} {\bibfield  {journal} {\bibinfo  {journal}
  {Phys. Rev. Lett.}\ }\textbf {\bibinfo {volume} {89}},\ \bibinfo {pages}
  {133602} (\bibinfo {year} {2002})}\BibitemShut {NoStop}%
\bibitem [{\citenamefont {Mok}\ \emph {et~al.}(2020)\citenamefont {Mok},
  \citenamefont {Bharti}, \citenamefont {Kwek},\ and\ \citenamefont
  {Bayat}}]{Bayat-2020}%
  \BibitemOpen
  \bibfield  {author} {\bibinfo {author} {\bibfnamefont {W.-K.}\ \bibnamefont
  {Mok}}, \bibinfo {author} {\bibfnamefont {K.}~\bibnamefont {Bharti}},
  \bibinfo {author} {\bibfnamefont {L.-C.}\ \bibnamefont {Kwek}}, \ and\
  \bibinfo {author} {\bibfnamefont {A.}~\bibnamefont {Bayat}},\ }\href@noop {}
  {\enquote {\bibinfo {title} {Optimal probes for global quantum
  thermometry},}\ } (\bibinfo {year} {2020}),\ \Eprint
  {http://arxiv.org/abs/2010.14200} {arXiv:2010.14200 [quant-ph]} \BibitemShut
  {NoStop}%
\bibitem [{\citenamefont {Rubio}\ \emph {et~al.}(2020)\citenamefont {Rubio},
  \citenamefont {Anders},\ and\ \citenamefont {Correa}}]{Correa-2020}%
  \BibitemOpen
  \bibfield  {author} {\bibinfo {author} {\bibfnamefont {J.}~\bibnamefont
  {Rubio}}, \bibinfo {author} {\bibfnamefont {J.}~\bibnamefont {Anders}}, \
  and\ \bibinfo {author} {\bibfnamefont {L.~A.}\ \bibnamefont {Correa}},\
  }\href@noop {} {\enquote {\bibinfo {title} {Global quantum thermometry},}\ }
  (\bibinfo {year} {2020}),\ \Eprint {http://arxiv.org/abs/2011.13018}
  {arXiv:2011.13018 [quant-ph]} \BibitemShut {NoStop}%
\bibitem [{\citenamefont {Bonfim}\ \emph {et~al.}(2019)\citenamefont {Bonfim},
  \citenamefont {Boechat},\ and\ \citenamefont {Florencio}}]{Bonfim-2019}%
  \BibitemOpen
  \bibfield  {author} {\bibinfo {author} {\bibfnamefont {O.~F. d.~A.}\
  \bibnamefont {Bonfim}}, \bibinfo {author} {\bibfnamefont {B.}~\bibnamefont
  {Boechat}}, \ and\ \bibinfo {author} {\bibfnamefont {J.}~\bibnamefont
  {Florencio}},\ }\href {\doibase 10.1103/PhysRevE.99.012122} {\bibfield
  {journal} {\bibinfo  {journal} {Phys. Rev. E}\ }\textbf {\bibinfo {volume}
  {99}},\ \bibinfo {pages} {012122} (\bibinfo {year} {2019})}\BibitemShut
  {NoStop}%
\bibitem [{\citenamefont {Cramer}(1946)}]{Cramer-1946}%
  \BibitemOpen
  \bibfield  {author} {\bibinfo {author} {\bibfnamefont {H.}~\bibnamefont
  {Cramer}},\ }\href@noop {} {\emph {\bibinfo {title} {Mathematical methods of
  statistics}}}\ (\bibinfo  {publisher} {Princeton University Press
  Princeton},\ \bibinfo {year} {1946})\BibitemShut {NoStop}%
\bibitem [{\citenamefont {Braunstein}\ and\ \citenamefont
  {Caves}(1994)}]{Braunstein-1994}%
  \BibitemOpen
  \bibfield  {author} {\bibinfo {author} {\bibfnamefont {S.~L.}\ \bibnamefont
  {Braunstein}}\ and\ \bibinfo {author} {\bibfnamefont {C.~M.}\ \bibnamefont
  {Caves}},\ }\href {\doibase 10.1103/PhysRevLett.72.3439} {\bibfield
  {journal} {\bibinfo  {journal} {Phys. Rev. Lett.}\ }\textbf {\bibinfo
  {volume} {72}},\ \bibinfo {pages} {3439} (\bibinfo {year}
  {1994})}\BibitemShut {NoStop}%
\bibitem [{\citenamefont {Paris}(2009)}]{Paris-2009}%
  \BibitemOpen
  \bibfield  {author} {\bibinfo {author} {\bibfnamefont {M.~G.~A.}\
  \bibnamefont {Paris}},\ }\href {\doibase 10.1142/S0219749909004839}
  {\bibfield  {journal} {\bibinfo  {journal} {International Journal of Quantum
  Information}\ }\textbf {\bibinfo {volume} {07}},\ \bibinfo {pages} {125}
  (\bibinfo {year} {2009})}\BibitemShut {NoStop}%
\bibitem [{\citenamefont {Liu}\ \emph {et~al.}(2019)\citenamefont {Liu},
  \citenamefont {Yuan}, \citenamefont {Lu},\ and\ \citenamefont
  {Wang}}]{Liu-2019}%
  \BibitemOpen
  \bibfield  {author} {\bibinfo {author} {\bibfnamefont {J.}~\bibnamefont
  {Liu}}, \bibinfo {author} {\bibfnamefont {H.}~\bibnamefont {Yuan}}, \bibinfo
  {author} {\bibfnamefont {X.-M.}\ \bibnamefont {Lu}}, \ and\ \bibinfo {author}
  {\bibfnamefont {X.}~\bibnamefont {Wang}},\ }\href {\doibase
  10.1088/1751-8121/ab5d4d} {\bibfield  {journal} {\bibinfo  {journal} {Journal
  of Physics A: Mathematical and Theoretical}\ }\textbf {\bibinfo {volume}
  {53}},\ \bibinfo {pages} {023001} (\bibinfo {year} {2019})}\BibitemShut
  {NoStop}%
\bibitem [{\citenamefont {Seveso}\ and\ \citenamefont
  {Paris}(2020)}]{Seveso-2020}%
  \BibitemOpen
  \bibfield  {author} {\bibinfo {author} {\bibfnamefont {L.}~\bibnamefont
  {Seveso}}\ and\ \bibinfo {author} {\bibfnamefont {M.~G.~A.}\ \bibnamefont
  {Paris}},\ }\href {\doibase 10.1142/S0219749920300016} {\bibfield  {journal}
  {\bibinfo  {journal} {International Journal of Quantum Information}\ }\textbf
  {\bibinfo {volume} {18}},\ \bibinfo {pages} {2030001} (\bibinfo {year}
  {2020})}\BibitemShut {NoStop}%
\bibitem [{\citenamefont {Albarelli}\ \emph {et~al.}(2020)\citenamefont
  {Albarelli}, \citenamefont {Barbieri}, \citenamefont {Genoni},\ and\
  \citenamefont {Gianani}}]{Albarelli-2020}%
  \BibitemOpen
  \bibfield  {author} {\bibinfo {author} {\bibfnamefont {F.}~\bibnamefont
  {Albarelli}}, \bibinfo {author} {\bibfnamefont {M.}~\bibnamefont {Barbieri}},
  \bibinfo {author} {\bibfnamefont {M.}~\bibnamefont {Genoni}}, \ and\ \bibinfo
  {author} {\bibfnamefont {I.}~\bibnamefont {Gianani}},\ }\href {\doibase
  https://doi.org/10.1016/j.physleta.2020.126311} {\bibfield  {journal}
  {\bibinfo  {journal} {Physics Letters A}\ }\textbf {\bibinfo {volume}
  {384}},\ \bibinfo {pages} {126311} (\bibinfo {year} {2020})}\BibitemShut
  {NoStop}%
\bibitem [{\citenamefont {Le~Cam}(1986)}]{LeCam-1986}%
  \BibitemOpen
  \bibfield  {author} {\bibinfo {author} {\bibfnamefont {L.~M.}\ \bibnamefont
  {Le~Cam}},\ }\href@noop {} {\emph {\bibinfo {title} {Asymptotic methods in
  statistical decision theory}}},\ Springer series in statistics\ (\bibinfo
  {publisher} {Springer-Verlag},\ \bibinfo {address} {New York},\ \bibinfo
  {year} {1986})\BibitemShut {NoStop}%
\bibitem [{\citenamefont {Hradil}\ \emph {et~al.}(1996)\citenamefont {Hradil},
  \citenamefont {My\ifmmode~\check{s}\else \v{s}\fi{}ka}, \citenamefont
  {Pe\ifmmode~\check{r}\else \v{r}\fi{}ina}, \citenamefont {Zawisky},
  \citenamefont {Hasegawa},\ and\ \citenamefont {Rauch}}]{Hradil-1996}%
  \BibitemOpen
  \bibfield  {author} {\bibinfo {author} {\bibfnamefont {Z.}~\bibnamefont
  {Hradil}}, \bibinfo {author} {\bibfnamefont {R.}~\bibnamefont
  {My\ifmmode~\check{s}\else \v{s}\fi{}ka}}, \bibinfo {author} {\bibfnamefont
  {J.}~\bibnamefont {Pe\ifmmode~\check{r}\else \v{r}\fi{}ina}}, \bibinfo
  {author} {\bibfnamefont {M.}~\bibnamefont {Zawisky}}, \bibinfo {author}
  {\bibfnamefont {Y.}~\bibnamefont {Hasegawa}}, \ and\ \bibinfo {author}
  {\bibfnamefont {H.}~\bibnamefont {Rauch}},\ }\href {\doibase
  10.1103/PhysRevLett.76.4295} {\bibfield  {journal} {\bibinfo  {journal}
  {Phys. Rev. Lett.}\ }\textbf {\bibinfo {volume} {76}},\ \bibinfo {pages}
  {4295} (\bibinfo {year} {1996})}\BibitemShut {NoStop}%
\bibitem [{\citenamefont {Pezz\'e}\ \emph {et~al.}(2007)\citenamefont
  {Pezz\'e}, \citenamefont {Smerzi}, \citenamefont {Khoury}, \citenamefont
  {Hodelin},\ and\ \citenamefont {Bouwmeester}}]{Pezze-2007}%
  \BibitemOpen
  \bibfield  {author} {\bibinfo {author} {\bibfnamefont {L.}~\bibnamefont
  {Pezz\'e}}, \bibinfo {author} {\bibfnamefont {A.}~\bibnamefont {Smerzi}},
  \bibinfo {author} {\bibfnamefont {G.}~\bibnamefont {Khoury}}, \bibinfo
  {author} {\bibfnamefont {J.~F.}\ \bibnamefont {Hodelin}}, \ and\ \bibinfo
  {author} {\bibfnamefont {D.}~\bibnamefont {Bouwmeester}},\ }\href {\doibase
  10.1103/PhysRevLett.99.223602} {\bibfield  {journal} {\bibinfo  {journal}
  {Phys. Rev. Lett.}\ }\textbf {\bibinfo {volume} {99}},\ \bibinfo {pages}
  {223602} (\bibinfo {year} {2007})}\BibitemShut {NoStop}%
\bibitem [{\citenamefont {Rubio}\ and\ \citenamefont
  {Dunningham}(2019)}]{Rubio-2019}%
  \BibitemOpen
  \bibfield  {author} {\bibinfo {author} {\bibfnamefont {J.}~\bibnamefont
  {Rubio}}\ and\ \bibinfo {author} {\bibfnamefont {J.}~\bibnamefont
  {Dunningham}},\ }\href {\doibase 10.1088/1367-2630/ab098b} {\bibfield
  {journal} {\bibinfo  {journal} {New Journal of Physics}\ }\textbf {\bibinfo
  {volume} {21}},\ \bibinfo {pages} {043037} (\bibinfo {year}
  {2019})}\BibitemShut {NoStop}%
\bibitem [{\citenamefont {Olivares}\ and\ \citenamefont
  {Paris}(2009)}]{Olivares-2009}%
  \BibitemOpen
  \bibfield  {author} {\bibinfo {author} {\bibfnamefont {S.}~\bibnamefont
  {Olivares}}\ and\ \bibinfo {author} {\bibfnamefont {M.~G.~A.}\ \bibnamefont
  {Paris}},\ }\href {\doibase 10.1088/0953-4075/42/5/055506} {\bibfield
  {journal} {\bibinfo  {journal} {Journal of Physics B: Atomic, Molecular and
  Optical Physics}\ }\textbf {\bibinfo {volume} {42}},\ \bibinfo {pages}
  {055506} (\bibinfo {year} {2009})}\BibitemShut {NoStop}%
\bibitem [{\citenamefont {Campos~Venuti}\ and\ \citenamefont
  {Zanardi}(2007)}]{Venuti-2007}%
  \BibitemOpen
  \bibfield  {author} {\bibinfo {author} {\bibfnamefont {L.}~\bibnamefont
  {Campos~Venuti}}\ and\ \bibinfo {author} {\bibfnamefont {P.}~\bibnamefont
  {Zanardi}},\ }\href {\doibase 10.1103/PhysRevLett.99.095701} {\bibfield
  {journal} {\bibinfo  {journal} {Phys. Rev. Lett.}\ }\textbf {\bibinfo
  {volume} {99}},\ \bibinfo {pages} {095701} (\bibinfo {year}
  {2007})}\BibitemShut {NoStop}%
\bibitem [{\citenamefont {Zanardi}\ and\ \citenamefont
  {Paunkovi\ifmmode~\acute{c}\else \'{c}\fi{}}(2006)}]{Zanardi-2006}%
  \BibitemOpen
  \bibfield  {author} {\bibinfo {author} {\bibfnamefont {P.}~\bibnamefont
  {Zanardi}}\ and\ \bibinfo {author} {\bibfnamefont {N.}~\bibnamefont
  {Paunkovi\ifmmode~\acute{c}\else \'{c}\fi{}}},\ }\href {\doibase
  10.1103/PhysRevE.74.031123} {\bibfield  {journal} {\bibinfo  {journal} {Phys.
  Rev. E}\ }\textbf {\bibinfo {volume} {74}},\ \bibinfo {pages} {031123}
  (\bibinfo {year} {2006})}\BibitemShut {NoStop}%
\bibitem [{\citenamefont {Zhou}\ \emph {et~al.}(2008)\citenamefont {Zhou},
  \citenamefont {Zhao},\ and\ \citenamefont {Li}}]{Zhou-2008}%
  \BibitemOpen
  \bibfield  {author} {\bibinfo {author} {\bibfnamefont {H.-Q.}\ \bibnamefont
  {Zhou}}, \bibinfo {author} {\bibfnamefont {J.-H.}\ \bibnamefont {Zhao}}, \
  and\ \bibinfo {author} {\bibfnamefont {B.}~\bibnamefont {Li}},\ }\href
  {\doibase 10.1088/1751-8113/41/49/492002} {\bibfield  {journal} {\bibinfo
  {journal} {Journal of Physics A: Mathematical and Theoretical}\ }\textbf
  {\bibinfo {volume} {41}},\ \bibinfo {pages} {492002} (\bibinfo {year}
  {2008})}\BibitemShut {NoStop}%
\bibitem [{\citenamefont {Zanardi}\ \emph {et~al.}(2007)\citenamefont
  {Zanardi}, \citenamefont {Giorda},\ and\ \citenamefont
  {Cozzini}}]{Zanardi-2007-crit}%
  \BibitemOpen
  \bibfield  {author} {\bibinfo {author} {\bibfnamefont {P.}~\bibnamefont
  {Zanardi}}, \bibinfo {author} {\bibfnamefont {P.}~\bibnamefont {Giorda}}, \
  and\ \bibinfo {author} {\bibfnamefont {M.}~\bibnamefont {Cozzini}},\ }\href
  {\doibase 10.1103/PhysRevLett.99.100603} {\bibfield  {journal} {\bibinfo
  {journal} {Phys. Rev. Lett.}\ }\textbf {\bibinfo {volume} {99}},\ \bibinfo
  {pages} {100603} (\bibinfo {year} {2007})}\BibitemShut {NoStop}%
\bibitem [{\citenamefont {Rezakhani}\ \emph {et~al.}(2010)\citenamefont
  {Rezakhani}, \citenamefont {Abasto}, \citenamefont {Lidar},\ and\
  \citenamefont {Zanardi}}]{Rezakhani-2010}%
  \BibitemOpen
  \bibfield  {author} {\bibinfo {author} {\bibfnamefont {A.~T.}\ \bibnamefont
  {Rezakhani}}, \bibinfo {author} {\bibfnamefont {D.~F.}\ \bibnamefont
  {Abasto}}, \bibinfo {author} {\bibfnamefont {D.~A.}\ \bibnamefont {Lidar}}, \
  and\ \bibinfo {author} {\bibfnamefont {P.}~\bibnamefont {Zanardi}},\ }\href
  {\doibase 10.1103/PhysRevA.82.012321} {\bibfield  {journal} {\bibinfo
  {journal} {Phys. Rev. A}\ }\textbf {\bibinfo {volume} {82}},\ \bibinfo
  {pages} {012321} (\bibinfo {year} {2010})}\BibitemShut {NoStop}%
\bibitem [{\citenamefont {Fisher}(1995)}]{Fisher-1995}%
  \BibitemOpen
  \bibfield  {author} {\bibinfo {author} {\bibfnamefont {D.~S.}\ \bibnamefont
  {Fisher}},\ }\href {\doibase 10.1103/PhysRevB.51.6411} {\bibfield  {journal}
  {\bibinfo  {journal} {Phys. Rev. B}\ }\textbf {\bibinfo {volume} {51}},\
  \bibinfo {pages} {6411} (\bibinfo {year} {1995})}\BibitemShut {NoStop}%
\bibitem [{\citenamefont {Sachdev}(2011)}]{Sachdev-2011}%
  \BibitemOpen
  \bibfield  {author} {\bibinfo {author} {\bibfnamefont {S.}~\bibnamefont
  {Sachdev}},\ }\href {\doibase 10.1017/CBO9780511973765} {\emph {\bibinfo
  {title} {Quantum Phase Transitions}}},\ \bibinfo {edition} {2nd}\ ed.\
  (\bibinfo  {publisher} {Cambridge University Press},\ \bibinfo {year}
  {2011})\BibitemShut {NoStop}%
\bibitem [{\citenamefont {Kokail}\ \emph {et~al.}(2019)\citenamefont {Kokail},
  \citenamefont {Maier}, \citenamefont {van Bijnen}, \citenamefont {Brydges},
  \citenamefont {Joshi}, \citenamefont {Jurcevic}, \citenamefont {Muschik},
  \citenamefont {Silvi}, \citenamefont {Blatt}, \citenamefont {Roos},\ and\
  \citenamefont {Zoller}}]{Kokail-2019}%
  \BibitemOpen
  \bibfield  {author} {\bibinfo {author} {\bibfnamefont {C.}~\bibnamefont
  {Kokail}}, \bibinfo {author} {\bibfnamefont {C.}~\bibnamefont {Maier}},
  \bibinfo {author} {\bibfnamefont {R.}~\bibnamefont {van Bijnen}}, \bibinfo
  {author} {\bibfnamefont {T.}~\bibnamefont {Brydges}}, \bibinfo {author}
  {\bibfnamefont {M.~K.}\ \bibnamefont {Joshi}}, \bibinfo {author}
  {\bibfnamefont {P.}~\bibnamefont {Jurcevic}}, \bibinfo {author}
  {\bibfnamefont {C.~A.}\ \bibnamefont {Muschik}}, \bibinfo {author}
  {\bibfnamefont {P.}~\bibnamefont {Silvi}}, \bibinfo {author} {\bibfnamefont
  {R.}~\bibnamefont {Blatt}}, \bibinfo {author} {\bibfnamefont {C.~F.}\
  \bibnamefont {Roos}}, \ and\ \bibinfo {author} {\bibfnamefont
  {P.}~\bibnamefont {Zoller}},\ }\href {\doibase 10.1038/s41586-019-1177-4}
  {\bibfield  {journal} {\bibinfo  {journal} {Nature}\ }\textbf {\bibinfo
  {volume} {569}},\ \bibinfo {pages} {355} (\bibinfo {year}
  {2019})}\BibitemShut {NoStop}%
\bibitem [{\citenamefont {Chiaro}\ \emph {et~al.}(2020)\citenamefont {Chiaro},
  \citenamefont {Neill}, \citenamefont {Bohrdt}, \citenamefont {Filippone},
  \citenamefont {Arute}, \citenamefont {Arya}, \citenamefont {Babbush},
  \citenamefont {Bacon}, \citenamefont {Bardin}, \citenamefont {Barends},
  \citenamefont {Boixo},\ and\ \citenamefont {et~al.}}]{Chiaro-2020}%
  \BibitemOpen
  \bibfield  {author} {\bibinfo {author} {\bibfnamefont {B.}~\bibnamefont
  {Chiaro}}, \bibinfo {author} {\bibfnamefont {C.}~\bibnamefont {Neill}},
  \bibinfo {author} {\bibfnamefont {A.}~\bibnamefont {Bohrdt}}, \bibinfo
  {author} {\bibfnamefont {M.}~\bibnamefont {Filippone}}, \bibinfo {author}
  {\bibfnamefont {F.}~\bibnamefont {Arute}}, \bibinfo {author} {\bibfnamefont
  {K.}~\bibnamefont {Arya}}, \bibinfo {author} {\bibfnamefont {R.}~\bibnamefont
  {Babbush}}, \bibinfo {author} {\bibfnamefont {D.}~\bibnamefont {Bacon}},
  \bibinfo {author} {\bibfnamefont {J.}~\bibnamefont {Bardin}}, \bibinfo
  {author} {\bibfnamefont {R.}~\bibnamefont {Barends}}, \bibinfo {author}
  {\bibfnamefont {S.}~\bibnamefont {Boixo}}, \ and\ \bibinfo {author}
  {\bibnamefont {et~al.}},\ }\href@noop {} {} (\bibinfo {year} {2020}),\
  \Eprint {http://arxiv.org/abs/1910.06024} {arXiv:1910.06024
  [cond-mat.dis-nn]} \BibitemShut {NoStop}%
\bibitem [{\citenamefont {Arute}\ \emph {et~al.}(2019)\citenamefont {Arute},
  \citenamefont {Arya}, \citenamefont {Babbush}, \citenamefont {Bacon},
  \citenamefont {Bardin}, \citenamefont {Barends}, \citenamefont {Biswas},
  \citenamefont {Boixo}, \citenamefont {Brandao}, \citenamefont {Buell},
  \citenamefont {Burkett},\ and\ \citenamefont {et~al.}}]{Arute-2019}%
  \BibitemOpen
  \bibfield  {author} {\bibinfo {author} {\bibfnamefont {F.}~\bibnamefont
  {Arute}}, \bibinfo {author} {\bibfnamefont {K.}~\bibnamefont {Arya}},
  \bibinfo {author} {\bibfnamefont {R.}~\bibnamefont {Babbush}}, \bibinfo
  {author} {\bibfnamefont {D.}~\bibnamefont {Bacon}}, \bibinfo {author}
  {\bibfnamefont {J.~C.}\ \bibnamefont {Bardin}}, \bibinfo {author}
  {\bibfnamefont {R.}~\bibnamefont {Barends}}, \bibinfo {author} {\bibfnamefont
  {R.}~\bibnamefont {Biswas}}, \bibinfo {author} {\bibfnamefont
  {S.}~\bibnamefont {Boixo}}, \bibinfo {author} {\bibfnamefont {F.~G. S.~L.}\
  \bibnamefont {Brandao}}, \bibinfo {author} {\bibfnamefont {D.~A.}\
  \bibnamefont {Buell}}, \bibinfo {author} {\bibfnamefont {B.}~\bibnamefont
  {Burkett}}, \ and\ \bibinfo {author} {\bibnamefont {et~al.}},\ }\href
  {\doibase 10.1038/s41586-019-1666-5} {\bibfield  {journal} {\bibinfo
  {journal} {Nature}\ }\textbf {\bibinfo {volume} {574}},\ \bibinfo {pages}
  {505} (\bibinfo {year} {2019})}\BibitemShut {NoStop}%
\bibitem [{\citenamefont {Dutta}\ \emph {et~al.}(2015)\citenamefont {Dutta},
  \citenamefont {Aeppli}, \citenamefont {Chakrabarti}, \citenamefont
  {Divakaran}, \citenamefont {Rosenbaum},\ and\ \citenamefont
  {Sen}}]{Dutta-2015}%
  \BibitemOpen
  \bibfield  {author} {\bibinfo {author} {\bibfnamefont {A.}~\bibnamefont
  {Dutta}}, \bibinfo {author} {\bibfnamefont {G.}~\bibnamefont {Aeppli}},
  \bibinfo {author} {\bibfnamefont {B.~K.}\ \bibnamefont {Chakrabarti}},
  \bibinfo {author} {\bibfnamefont {U.}~\bibnamefont {Divakaran}}, \bibinfo
  {author} {\bibfnamefont {T.~F.}\ \bibnamefont {Rosenbaum}}, \ and\ \bibinfo
  {author} {\bibfnamefont {D.}~\bibnamefont {Sen}},\ }\href {\doibase
  10.1017/CBO9781107706057} {\emph {\bibinfo {title} {Quantum Phase Transitions
  in Transverse Field Spin Models: From Statistical Physics to Quantum
  Information}}}\ (\bibinfo  {publisher} {Cambridge University Press},\
  \bibinfo {year} {2015})\BibitemShut {NoStop}%
\bibitem [{\citenamefont {Damski}\ and\ \citenamefont
  {Rams}(2013)}]{Damski-2013}%
  \BibitemOpen
  \bibfield  {author} {\bibinfo {author} {\bibfnamefont {B.}~\bibnamefont
  {Damski}}\ and\ \bibinfo {author} {\bibfnamefont {M.~M.}\ \bibnamefont
  {Rams}},\ }\href {\doibase 10.1088/1751-8113/47/2/025303} {\bibfield
  {journal} {\bibinfo  {journal} {Journal of Physics A: Mathematical and
  Theoretical}\ }\textbf {\bibinfo {volume} {47}},\ \bibinfo {pages} {025303}
  (\bibinfo {year} {2013})}\BibitemShut {NoStop}%
\end{thebibliography}%


\clearpage
\onecolumngrid
\newpage
\widetext
\begin{center}
\textbf{\large Supplemental Material: Global sensing and its impact for quantum many-body probes with criticality}

\vspace{0.25cm}

Victor Montenegro$^{1}$, Utkarsh Mishra$^{1}$, and Abolfazl Bayat$^{1}$

\vspace{0.25cm}

$^{1}${\small \em Institute of Fundamental and Frontier Sciences,\\ University of Electronic Science and Technology of China, Chengdu 610051, PR China}


\end{center}
\date{\today}
\setcounter{equation}{0}
\setcounter{figure}{0}
\setcounter{table}{0}
\setcounter{page}{1}
\makeatletter
\renewcommand{\theequation}{S\arabic{equation}}
\renewcommand{\thefigure}{S\arabic{figure}}

In this section, we analytically derive the solution in obtaining the ground state $|\Phi\rangle$ of the Hamiltonian of Eq.~\eqref{eq:hamiltonian} in the absence of any longitudinal field present in the system. In addition, we will also show how to evaluate the QFI for this case. 

To derive the solution we consider the Hamiltonian
\begin{equation}
H = J\sum_{i=1}^{L} \sigma_x^i \sigma_x^{i+1} - \sum_{i=1}^{L} h_z\sigma_z^i,\label{eq:sm-hamiltonian}
\end{equation}
where, $\sigma_\alpha^i$ ($\alpha = x,z$) is the Pauli operator at site $i$, $J>0$ is the exchange interaction, $h_z$ is the transverse field, and periodic boundary condition is imposed. The ground state of the Hamiltonian in Eq.~(\ref{eq:sm-hamiltonian}) can be obtained by first writing $H$ in c-fermionic and then to Bogolyubov basis. Following the treatment of diagonalization of $H$ as in Ref.~\cite{Dutta-2015, Sachdev-2011, Damski-2013}, we define raising and lowering operator at a site $i$ by $\sigma_{\pm}^{i}=(\sigma_{i}^{x}\pm\sigma_{i}^{y})/2$.  The Hamiltonian in terms of spin lowering and raising operator then becomes 
 
 \begin{equation}
H = J\sum_{i=1}^{L} (\sigma_+^i \sigma_+^{i+1}+\sigma_-^i \sigma_-^{i+1}+\sigma_+^i \sigma_-^{i+1}+\sigma_-^i \sigma_+^{i+1}) - \sum_{i=1}^{L} h_z\sigma_z^i.
\end{equation}
The next step is to apply Jordan-Wigner transformation which transforms the Hamiltonian to c-fermionic basis. The Jordan-Wigner transformation is defined as $\sigma^{j}_{-}=\exp[-i\sum_{\ell=1}^{j-1}\sigma^{\ell}_{+}\sigma^{\ell}_{-}]c_{j}$ , $\sigma^{j}_{+}=c^{\dagger}_ {j}\exp[i\sum_{\ell=1}^{j-1}\sigma^{\ell}_{+}\sigma^{\ell}_{-}]$, and $\sigma^{i}_{z}=2c^{\dagger}_{i}c_{i}-\mathbb{I}$. 
The c-fermions follow the fermionic commutation relation as $\{c_{i},c^{\dagger}_{j}\}=\delta_{ij}$ and $\{c_{i},c_{j}\}=0$. Now by exploiting the translational invariance of the original Hamiltonian, the above Hamiltonian can be written in Fourier space as follow  
 \begin{equation}
H = \sum_{k>0} \begin{pmatrix}
  c^{\dagger}_{k} &c_{-k}
\end{pmatrix} \begin{pmatrix}
  h_z+J\cos(k) & i\sin(k)\\ 
  -i\sin(k) & -(h_z+J\cos(k)
\end{pmatrix}
\begin{pmatrix}
  c_{k} \\
  c_{-k}^{\dagger}.
\end{pmatrix} \label{eq:sm-FS-hamiltonian}
\end{equation}
In the above equation, we used $c^{\ell}=\frac{1}{\sqrt{N}}\sum_{k}c_{k}e^{ik\ell}$ and the basis of the matrix is $\{|0\rangle,|k,-k\rangle\}$ with $k=\pm\frac{\pi}{3},\pm \frac{\pi}{5}, \ldots,\pm\frac{(N-1)\pi}{N}$. Now we introduce a matrix $S = \begin{pmatrix}
  \cos(\theta_k/2) &e^{i\phi}\sin(\theta_k/2) \\
  e^{-i\phi}\sin(\theta_k/2)&\cos(\theta_k/2)\end{pmatrix}$ such that  $\begin{pmatrix}
  \gamma_{k} \\
  \gamma{\dagger}_{-k}.
\end{pmatrix}=S\begin{pmatrix}
  c_{k} \\
  c{\dagger}_{-k}.
\end{pmatrix}.$ 
By substituting $S$ in Eq.~\eqref{eq:sm-FS-hamiltonian}, we arrive at the diagonal form of the Hamiltonian expressed as 
 \begin{equation}
H = \sum_{k>0} \begin{pmatrix}
  \gamma^{\dagger}_{k} &\gamma_{-k}
\end{pmatrix} \begin{pmatrix}
  \epsilon_{k} & 0\\ 
  0 & -\epsilon_{k}
\end{pmatrix}
\begin{pmatrix}
  \gamma_{k} \\
  \gamma{\dagger}_{-k}.
\end{pmatrix}. \label{eq:diag-sm-FS-hamiltonian}
\end{equation}
The operators $\{\gamma^{\dagger}_{k},\gamma_{k}\}$ are known as the Bogoliubov operators. The ground state energy of the system is given by $E=-\sum_{k>0}\epsilon_{k}=-\sum_{k>0}\sqrt{(h_z+J\cos(k))^2+J^2\sin^{2}(k)}$. The ground state is annihilated  by all the operators $\gamma_{k}$ and can be written as
\begin{equation}
|\Phi\rangle = \Pi_{k}\Big(\cos(\theta_k/2)+\sin(\theta_k/2)c^{\dagger}_{k}c^{\dagger}_{-k}\Big)|vac\rangle, \label{eq:BCS_ground}   
\end{equation}
where, $|vac\rangle$ is annihilated by all $c_{k}$ operators. Furthermore, $\Big(\sin(\theta_k/2),\cos(\theta_k/2)\Big) = \Big(\frac{J\sin(k)}{\epsilon_k},\frac{h_z+J\cos(k)}{\epsilon_k}\Big)$.

For the above ground state with the tuning parameter $h_z$, we can define ground state fidelity as \begin{equation}
F = \langle \Phi (h_z)|\Phi(h_z+\Delta h_z)\rangle,
\end{equation}
where, $\Delta h_z$ is the small shift in $h_z$. By performing Taylor expansion for $\Delta h_z\ll 1$, we have
\begin{equation}
F = 1-\frac{1}{2}\chi (\Delta h_z)^2,  \label{eq:fidelity_sus} 
\end{equation}
where, we have used the normalization condition i.e., $\langle \Phi(h_z)|\Phi(h_z)\rangle=1$. In Eq.~(\ref{eq:fidelity_sus}), $\chi=-\langle \Phi(h_z)|\frac{\partial^2 |\Phi(h_z)}{\partial h^2_z}\rangle$ is called the fidelity susceptibility. In terms of fidelity, the fidelity susceptibility is given by $\chi = 2(1-F)/(\Delta h_z)^2$. The fidelity susceptibility is directly proportional to the ${\cal F}_{Q}$ and is given by ${\cal F}_{Q}=4\chi$. For the ground state in Eq.~(\ref{eq:BCS_ground}), the fidelity is given by $F=\sum_{k>0}\cos\left( \frac{\theta_{k}(h_z)-\theta_{k}(h_z+\Delta h_z)}{2}\right)$. We used this approach to calculate the quantum Fisher information in the main text for exactly solvable case. 
\begin{figure}[t]
\centering \includegraphics[scale=0.5]{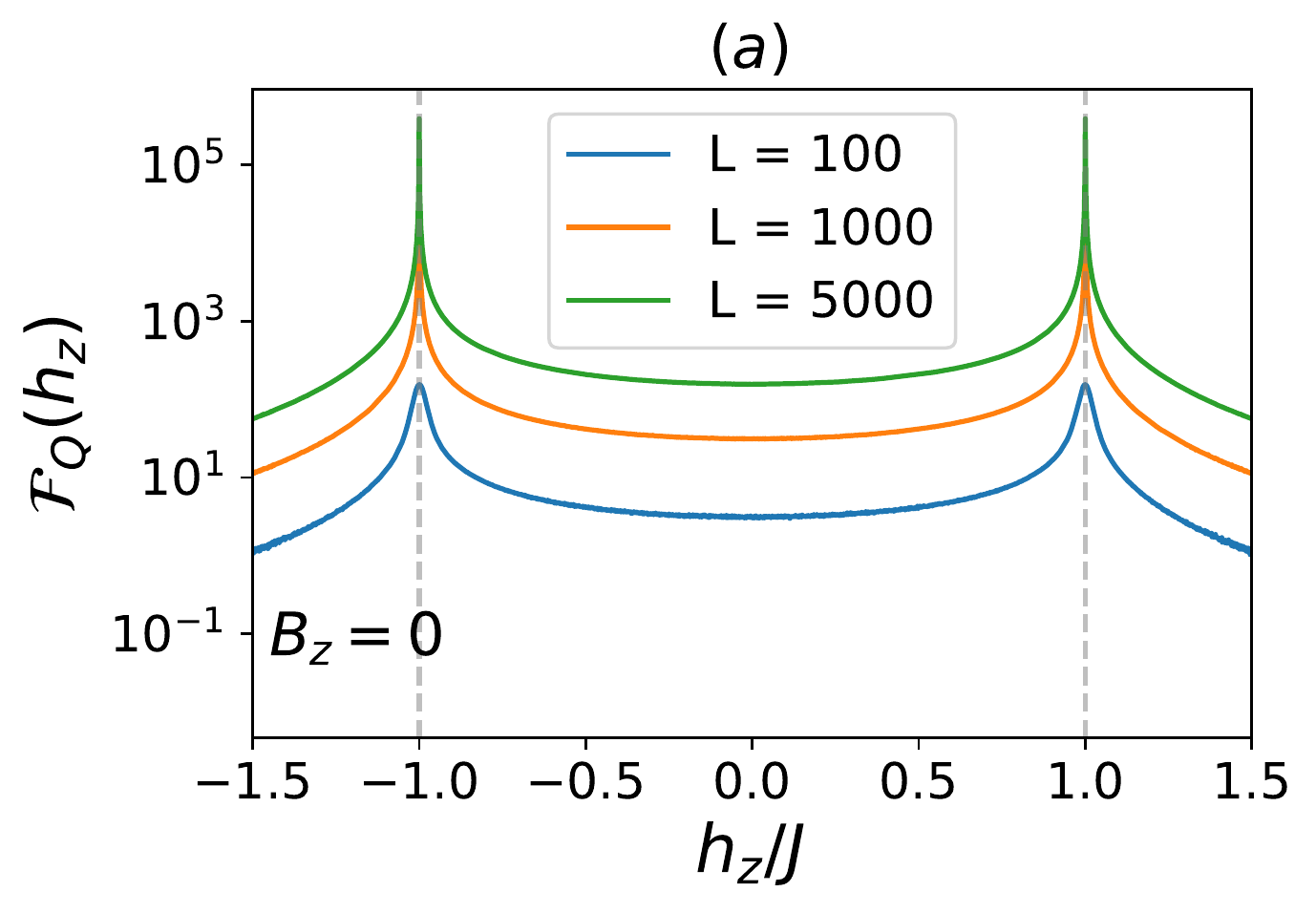}\includegraphics[scale=0.5]{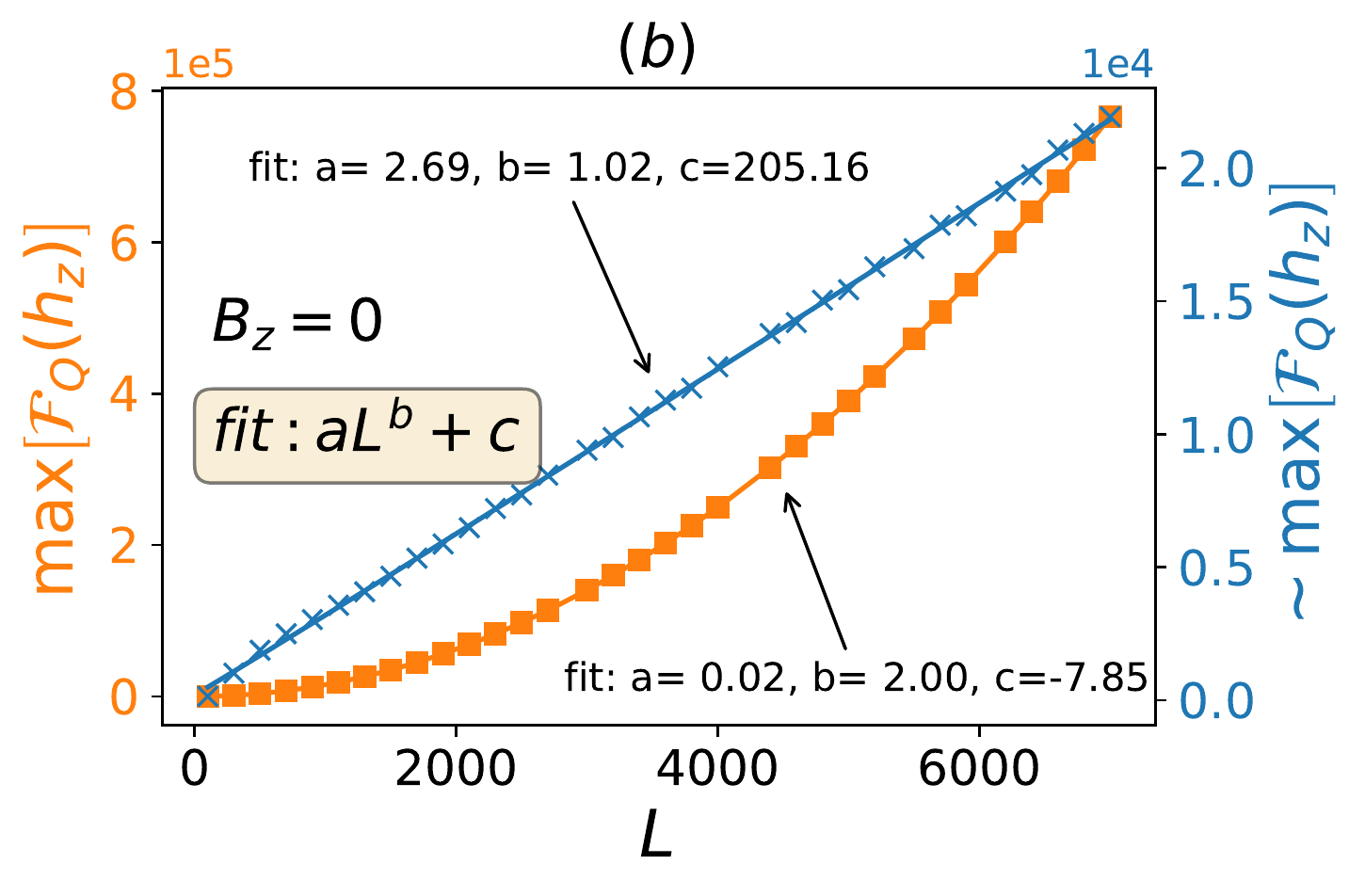}
\caption{(a) Quantum Fisher information $\mathcal{F}_Q(h_z)$ as a function of $h_z/J$ for various system lengths $L$. (b) Scaling of the quantum Fisher information with respect to the system size $L$ at the critical point $h = h^{\mathrm{crit}}$ (left $y-$ axis) and around its critical point $h \approx h^{\mathrm{crit}}$ (right $y-$axis).}\label{fig:sup-material}
\end{figure}

For the sake of completeness, we plot in Fig.~\ref{fig:sup-material}(a), the QFI $\mathcal{F}_Q(h_z)$ as a function of $h_z/J$ for different system sizes $L$ as computed using the above fidelity susceptibility approach. As evident from the figure, the QFI peaks at the critical point of the system $h_z = h_z^\mathrm{crit}$. As the system size increases, the QFI takes higher values. However, one can notice that this feature only occurs in a very narrow vicinity around its critical point. To determine how this narrow vicinity can diminish the sensing precision, we plot in Fig.~\ref{fig:sup-material}(b), the maximum of the QFI at the critical point (left $y-$axis) and around its vicinity (right $y-$axis) with their corresponding fitting curves $aL^{b} + c$. As the figure shows, at the critical point, the QFI scales as $\sim L^2$ as determined by the $b \approx 2$ fitting coefficient. A modest deviation of $h_z^\mathrm{crit}$, here $h_z^\mathrm{crit} + 0.005J$, makes the sensing precision to drop to its standard limit, i.e. $\sim L$.
\end{document}